\documentclass[11pt]{article}
\usepackage{amsmath, amssymb, amsthm}
\usepackage{graphicx}
\usepackage{hyperref}
\usepackage{geometry}
\usepackage{braket}
\usepackage{tikz}
\usetikzlibrary{quantikz2}
\usetikzlibrary{graphs, arrows.meta, positioning}
\usepackage{subcaption}
\usepackage{upgreek}
\usepackage{mathrsfs}
\usepackage{arydshln}
\usepackage{bm}
\usepackage{float}
\usetikzlibrary{backgrounds}

\title{Generalized Efficient Quantum Circuit Implementation of Discrete-Time Quantum Walks on Cayley Graphs}
\author{
  Seoyoon Kang\\
  Department of Physics, Worcester Polytechnic Institute,\\ Worcester, MA 01609, USA\\[0.75em]
}
\date{\today}

\begin{document}
\maketitle

\begin{abstract}
We present a generalized and efficient quantum circuit framework for implementing discrete-time quantum walks (DTQWs) on Cayley graphs of arbitrary dimension. Building on the Boundary QFT scheme of Razzoli et al., we introduce a systematic multi-stage decomposition of the shift operator for 1D Cayley graphs across three classes of generating sets: inverse-closed without involutions, inverse-closed with an involution, and non-inverse-closed. The decomposition hierarchically factorizes the QFT-diagonalized shift operator into structured block components, progressively reducing the control degree of the required rotation gates and replacing high-degree multi-qubit controlled operations with collections of lower-degree equivalents. We extend this construction to $d$-dimensional torus graphs and provide explicit circuit implementations for an 8-Cayley graph and a $\mathbb{Z}_{16} \times \mathbb{Z}_8$ torus graph as concrete illustrations. Gate complexity analysis using the linear CNOT scaling of Rosa et al. demonstrates that the decomposed implementation achieves a substantial reduction in upper-bound CNOT cost relative to the naive implementation within the regime $k \leq 64$ for inverse-closed graphs and $k \leq 16$ for non-inverse-closed graphs, where $k$ denotes the degree of the generating set. Benchmarking further reveals that this efficiency gain is largely insensitive to the system size $N$, identifying $k$ as the dominant resource parameter for the shift operator. These results provide a scalable and hardware-conscious pathway toward practical DTQW implementations on near-term quantum devices.
\end{abstract}

\section{Introduction}
Discrete-time quantum walks (DTQW) serves as the quantum analogue of classical random walk and constitute a universal model of quantum computation ~\cite{lovett_universal_2010, chawla_multi-qubit_2023, singh_universal_2021}. Their versatility has provided a robust framework for designing various quantum algorithms, including but not limited to quantum state transport \cite{kurzynski_discrete-time_2011, nitsche_quantum_2016}, quantum search \cite{childs_spatial_2004, shenvi_quantum_2003}, and simulation of physical systems \cite{berry_black-box_2012, nejadsattari_experimental_2019, sansoni_two-particle_2012}. Quantum walks are realized with several distinct models, such as continuous-time quantum walks implemented via Hamiltonian encoding of a graph's adjacency matrix \cite{farhi_quantum_1998}, staggered quantum walks using graph tessellations \cite{portugal_staggered_2016}, Szegedy Quantum Walk based on classical Markov chains \cite{szegedy_quantum_2004}, and discrete-time quantum walk \cite{aharonov_quantum_1993}, which this work focuses on. Discrete-time quantum walk is particularly suited for practical circuit implementation due to its discretized nature. Various studies have explored circuit-level realization of DTQWs \cite{wing-bocanegra_circuit_2023, wing-bocanegra_unitary_2023, olivieri_experimental_2021, sarkar_quantum_2024}. 

In the DTQW framework, the walker evolves through discrete steps comprising (i) a coin operation and (ii) a conditional shift operation, each defined with its respective operator. The coin operator acts on an internal degree of freedom to create a coherent superposition of possible directions, analogous to classical coin toss. The shift operator, acting on the entire Hilbert space, then updates the walker's position conditioned on the state of the coin. 

For circuit implementations, the efficiency of a DTQW is determined by the effective compilation of these two operators. While the coin operator $C$, often a structured unitary such as a Hadamard or higher-dimensional rotation, typically requires constant resources for graphs of a fixed degree, the shift operator $S$ implements controlled modular translations on the position register. As the system size scales, the realization of the walk hinges largely on decomposing $S$ into gate-efficient circuits.  Because the coin operator remains relatively "cheap" compared to the shift operator, this work focuses exclusively on the efficient circuit implementation of $S$.

While DTQWs on simple cycle graphs admit relatively straightforward circuit implementations, extending these constructions to general 1D Cayley graphs and higher dimensions introduces significant overhead in multi-qubit controlled operations. Such overhead is undesirable in Noisy Intermediate-scale Quantum (NISQ) devices, which are constrained by decoherence and high error rates in two-qubit gates that limit the circuit depth and size, and specifically the complexity of two-qubit gates due to its high error rate. These limitations have prompted the development of implementation methods that target reductions in circuit depth, size, and control degree.

Earlier approaches relied on the direct implementation of the shift operator, leading to high control degrees. 
Douglas et al. \cite{douglas_efficient_2009} provided the first explicit circuit implementations of DTQW on various graphs using series of controlled gates. Shakeel \cite{shakeel_efficient_2020} improved efficiency by utilizing the Quantum Fourier Transform (QFT) to diagonalize the shift operation. Most recently, Razzoli et al. \cite{razzoli_efficient_2024} proposed a systematic factorization of the shift operator that achieves the state-of-the-art circuit performance, and significantly reduces two-qubit gate counts by eliminating intermediate QFTs at each time step.

In this work, we extend the method proposed in Razzoli et al. and present a systematic, multi-stage decomposition method of the shift operator for Cayley graphs with different generating sets and of arbitrary dimensions. Our approach factorizes the QFT-diagonalized operator into structured block components, enabling binary-decomposition-based control optimization. This reduces the degree of multi-controlled rotation gates and, consequently, the number of elementary CNOT gates required for their implementation \cite{rosa_optimizing_2025}. We compare the multi-qubit controlled gate complexities and the upper-bound CNOT costs of our decomposed implementation against the naive implementation (without any decomposition of the shift operator) introduced in Douglas et al. \cite{douglas_efficient_2009} across several Cayley graph examples. By applying the CNOT scaling established in Rosa et al. \cite{rosa_optimizing_2025}, we demonstrate that our method yields a substantial reduction in the total count of elementary two-qubit gates under a certain regime, $k\leq64$, where $k$ is the degree of a generating set. 

The remainder of this paper is organized as follows. In Section~\ref{Preliminaries}, we review the fundamental definitions of Cayley graph and the existing DTQW implementation schemes upon which this work builds. Section~\ref{1D_cayley_graph} introduces our decomposition method of the shift operator in the context of 1D Cayley graphs, covering three types of generating sets: (i) inverse-closed without involutions, (ii) inverse-closed with an involution, and (iii) non-inverse-closed. We also provide a concrete circuit implementation for an $8$-Cayley graph, followed by benchmarking results that compares the upper-bound CNOT cost scaling before and after the decomposition. Section~\ref{2D_cayley_graph} extends this decomposition to 2D and arbitrary dimensions, including an implementation for a 2D torus grid graph. Finally, Section~\ref{Conclusion} summarizes the advantages of our method and discusses directions for future work. 


\section{Preliminaries}
\label{Preliminaries}
In this section, we review the fundamental definitions of Cayley graphs and establish the primary variables utilized throughout this discussion. We also summarize the historical development of quantum circuit implementation methods for DTQW on an N-cycle graph. This survey encompasses the Increment/Decrement (ID) scheme introduced by Douglas et al.~\cite{douglas_efficient_2009}, the Stepwise QFT scheme proposed by Shakeel et al.~\cite{shakeel_efficient_2020}, and the Boundary QFT scheme detailed by Razzoli et al.~\cite{razzoli_efficient_2024}, the latter of which we generalize to Cayley graphs in this work. 

\subsection{Cayley Graph}
Given a group G and a generating set $\mathcal{S}$, a Cayley graph $\Gamma(G, \mathcal{S})$ is defined such that each vertex corresponds to an element $g \in G$, and directed edges are constructed from $g$ to $g + \sigma$ for every generator $\sigma \in \mathcal{S}$. When the generating set is symmetric ($\mathcal{S} = \mathcal{S}^{-1}$) or inverse-closed, that is, for every generator $\sigma \in \mathcal{S}$, its inverse $\sigma^{-1}$ is also in $\mathcal{S}$, the graph is undirected. Within such a symmetric generating set, an involution may exist---an element that is its own inverse ($\sigma = \sigma^{-1}$). If the generating set $\mathcal{S}$ is not symmetric—--that is, it is not closed under inversion---the graph is directed. Examples of undirected and directed Cayley graphs with and without involutions are illustrated in Fig.~\ref{fig:cayley_graphs} below. Cayley graphs are inherently $k$-regular, with $k = |\mathcal{S}|$, and exhibit vertex-transitivity, reflecting the underlying symmetry of the group. For instance, the cyclic group $\mathbb{Z}_N$ with $\mathcal{S}=\{1, -1, 2, -2 \}$ produces an undirected $4$-regular graph where each vertex is connected to its neighbors modulo $N$. Similarly, the finite abelian group $\mathbb{Z}_N \times \mathbb{Z}_N$ with $\mathcal{S}=\{(1,0),(0,1)\}$ corresponds to an $N \times N$ toroidal lattice. 


\begin{figure}[h!]
    \centering
    \begin{subfigure}{0.45\textwidth}
    \centering
    \begin{tikzpicture}[every node/.style={circle, draw, fill=gray!30, minimum size=6mm}]
        \def\n{6}
        \def\r{2.5cm}
        \def\startAngle{90}
        \def\offset{pt}
    
        \foreach \i in {1,...,\n} {
            \pgfmathtruncatemacro{\label}{\i-1} 
            \node (N\i) at ({\startAngle - 360/\n * (\i-1)}:\r) {$\label$};
        }
    
        \foreach \i in {1,...,\n} {
            \pgfmathtruncatemacro{\next}{mod(\i,\n)+1}
            \draw[->, >=Stealth, line width=0.7pt, bend left=7] (N\i) to (N\next);
            \draw[<-, red, line width=0.7pt, >=Stealth, bend right=7] (N\i) to (N\next);
        }
        \foreach \i in {1,...,\n} {
            \pgfmathtruncatemacro{\target}{mod(\i+1,\n)+1} 
            \draw[->, blue, line width=0.7pt, >=Stealth, bend left=7] (N\i) to (N\target);
            \draw[<-, green!50!black, line width=0.7pt, >=Stealth, bend right=7] (N\i) to (N\target);
        }
        \foreach \i in {1,...,\n} {
            \pgfmathtruncatemacro{\target}{mod(\i+2,\n)+1} 
            \draw[->, orange, line width=0.7pt, >=Stealth, bend left=0] (N\i) to (N\target);
        }
    \end{tikzpicture}
    \caption*{$\Gamma(\mathbb{Z}_6, \{1, \textcolor{red}{-1}, \textcolor{blue}{2}, \textcolor{green!50!black}{-2}, \textcolor{orange}{3}\})$}
    \caption{}
    \end{subfigure}
    \hfill
    \begin{subfigure}{0.45\textwidth}
    \centering
    \begin{tikzpicture}[every node/.style={circle, draw, fill=gray!30, minimum size=3mm}]
        \def\n{3} 
        \def\spacing{2cm}
    
        \foreach \x in {0,...,2} {
            \foreach \y in {0,...,2} {
                \node (N\x\y) at (\y*\spacing, -\x*\spacing) {(\x,\y)};
            }
        }
    
        \foreach \x in {0,...,2} {
            \draw[->, blue, line width=0.5pt, >=Stealth] (N\x0) to (N\x1);
            \draw[->, blue, line width=0.5pt, >=Stealth] (N\x1) to (N\x2);     
            \draw[->, blue, line width=0.5pt, >=Stealth, bend right=30] (N\x2) to (N\x0);  
        }
        \foreach \y in {0,...,2} {
            \draw[->, red, line width=0.5pt, >=Stealth] (N0\y) to (N1\y);
            \draw[->, red, line width=0.5pt, >=Stealth] (N1\y) to (N2\y);  
            \draw[->, red, line width=0.5pt, >=Stealth, bend right=30] (N2\y) to (N0\y);  
        }
       
    \end{tikzpicture}
    \caption*{$\Gamma(\textcolor{red}{\mathbb{Z}_3} \times \textcolor{blue}{\mathbb{Z}_3}, \{\textcolor{red}{(1,0)}, \textcolor{blue}{(0,1)}\})$}
    \caption{}
    \end{subfigure}
    \caption{Examples of different types of Cayley graphs. (a) An undirected regular graph with an involution at $\sigma=3$. Each edge is color-coded according to its corresponding generator in the generating set. (b) A directed torus grid graph with each cyclic group and its generator edges color-coded accordingly.}
    \label{fig:cayley_graphs}
\end{figure}
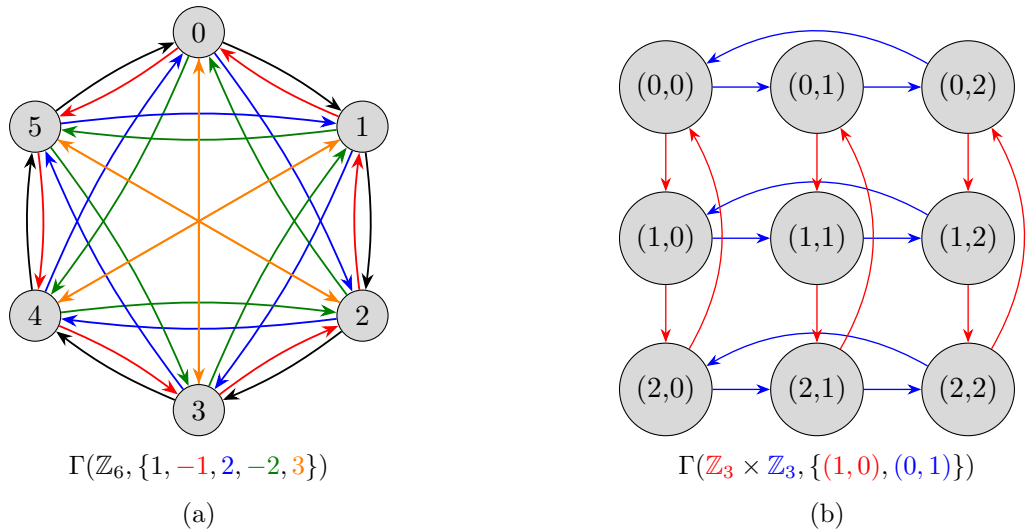

\subsection{DTQW Circuit Implementation Models}
We now introduce the quantum circuit setup for DTQW and review the evolution of implementation models for DTQW demonstrated on an $N$-cycle graph, a Cayley graph with a single generator and its inverse, $\Gamma(\mathbb{Z}_N, \{1,-1\})$. These models represent a sequential refinement of evolved implementation strategies rather than disjoint approaches; each subsequent section builds directly upon the concepts established in the preceding section.

\subsubsection{Quantum Circuit Setup}
\label{sec:circuit_setup}
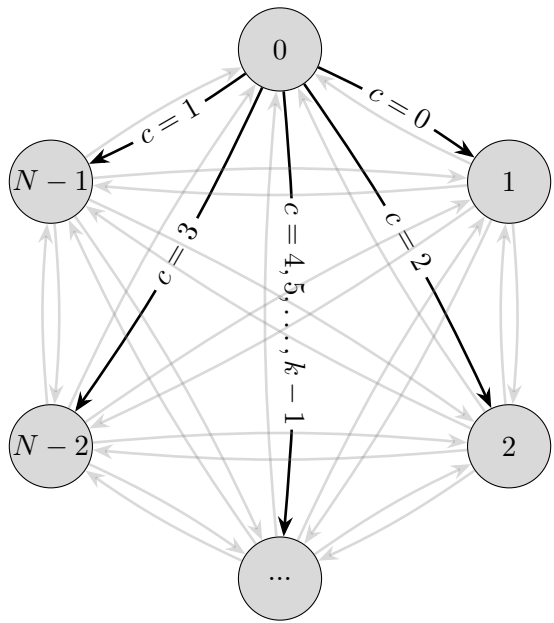
\begin{figure}[h!]
\centering
\begin{tikzpicture}[every node/.style={circle, draw, fill=gray!30, inner sep=0pt, minimum size=11mm}]
    \def\n{6}
    \def\r{3.5cm}
    \def\startAngle{90}
    \def\offset{pt}

    \node (N1) at ({\startAngle - 0}:\r) {$0$};
    \node (N2) at ({\startAngle - 360/6}:\r) {$1$};
    \node (N3) at ({\startAngle - 360/6*2}:\r) {$2$};
    \node (N4) at ({\startAngle - 360/6*3}:\r) {$...$};
    \node (N5) at ({\startAngle - 360/6*4}:\r) {$N-2$};
    \node (N6) at ({\startAngle - 360/6*5}:\r) {$N-1$};

    \draw[->, >=Stealth, line width=1pt, bend left=5] (N1) to node[pos=0.5, above=-15pt, sloped, inner sep=1pt, fill=white, draw=none] {$c=0$}  (N2);
    \draw[<-, >=Stealth, line width=1pt, bend right=5] (N6) to node[pos=0.5, below=-17pt, sloped, inner sep=1pt, fill=white, draw=none] {$c=1$} (N1);
    \draw[->, >=Stealth, line width=1pt, bend left=5] (N1) to node[pos=0.5, above=-15pt, sloped, inner sep=1pt, fill=white, draw=none] {$c=2$}  (N3);
    \draw[<-, >=Stealth, line width=1pt, bend right=5] (N5) to node[pos=0.5, below=-17pt, sloped, inner sep=1pt, fill=white, draw=none] {$c=3$} (N1);
    \draw[->, >=Stealth, line width=1pt, bend left=5] (N1) to node[pos=0.5, above=-45pt, sloped, inner sep=1pt, fill=white, draw=none] {$c=4,5,\dots,k-1$}  (N4);

    \foreach \i in {1,...,\n} {
        \pgfmathtruncatemacro{\next}{mod(\i,\n)+1}
        \pgfmathtruncatemacro{\nexttwo}{mod(\i+1,\n)+1}
        \pgfmathtruncatemacro{\nextthree}{mod(\i+2,\n)+1}
    
        \ifnum\i=1
        \else
            \draw[->, >=Stealth, line width=1pt, draw=gray, opacity=0.3, bend left=5](N\i) to (N\next);
    
            \draw[->, >=Stealth, line width=1pt, draw=gray, opacity=0.3, bend left=5](N\i) to (N\nexttwo);
    
            \draw[->, >=Stealth, line width=1pt, draw=gray, opacity=0.3, bend left=5](N\i) to (N\nextthree);
        \fi

        \ifnum\i=6
        \else
            \draw[<-, >=Stealth, line width=1pt, draw=gray, opacity=0.3, bend right=5](N\i) to (N\next);
        \fi

        \ifnum\i=5
        \else
            \draw[<-, >=Stealth, line width=1pt, draw=gray, opacity=0.3, bend right=5](N\i) to (N\nexttwo);
        \fi
    }
    
\end{tikzpicture}
\caption{Graphical representation of the DTQW setup on an $N$-node Cayley graph with $k$ generators. Each node represents a position state $|x\rangle$ in the Hilbert space $\mathcal{H}_p^{(N)}$. The directed edges originating from node $0$ illustrate the conditional shifts to neighboring nodes, where each transition is governed by a specific coin state $|c\rangle$ from the $k$-dimensional coin space $\mathcal{H}_c^{(k)}$.}
\label{fig:DTQW_setup}
\end{figure}

A quantum walker in DTQW is characterized by two degrees of freedom: the coin and the position. These are defined by the Hilbert spaces $\mathcal{H}_c^{(k)} = \text{span}(\{\ket{c} : c = 0,1,\dots, k-1\})$ and $\mathcal{H}_p^{(N)} = \text{span}(\{\ket{x} : x = 0,1, \dots, N-1\})$, respectively. Here, $c$ denotes the coin state associated with each generator defining the edge connections, and $x$ denotes the walker's position or node index. For a general Cayley graph without an involution, the generators $\sigma = 1, -1, \dots, k/2, -k/2$ are mapped to coin states $c=0, 1,\dots , k-2, k-1 $, as shown in Fig.~\ref{fig:DTQW_setup} \textcolor{red}. The full Hilbert Space is given by the tensor product $\mathcal{H}=\mathcal{H}_c\otimes \mathcal{H}_p$.

For a Cayley graph with $N=2^n$ vertices and a generating set of degree $k=2^\alpha$, a total of $n+\alpha$ qubits are required. For simplicity of the implementation, we restrict both $N$ and $k$ to be powers of two. In the case of an involution, where $k$ is odd, the coin register is padded with identity operations. The first $\alpha$ qubits comprise the coin register, while the remaining $n$ qubits form the position register. Following the little-endian convention, a general state of the walker is expressed as 
\begin{equation}
    \ket{\psi}_{walker} = \ket{c}\ket{x} = \ket{q^c_{\alpha}q^c_{\alpha-1}\dots q^c_0} \ket{q^p_nq^p_{n-1} \dots q^p_0}.
\end{equation}

The evolution of a single-step DTQW is governed by the unitary operator $U = S(C\otimes I_p)$, where $C$ is a coin operator that only acts on the coin register, superposing all coin states, and $S$ is a shift operator which acts on the full Hilbert space, shifting the position of the walker according to the coin state. The coin operator defines the amplitude associated with each coin state, and different choices of coin, such as Hadamard and Grover~\cite{ambainis_coins_2005}, lead to different interference patterns of the walk. 

On the other hand, the shift operator $S$ has a form of 
\begin{equation}
    S = \sum_{c=0}^{k-1} \sum_{x=0}^{N-1} \ket{c}\bra{c} \otimes \ket{[x + (-1)^{c}\,\bigr(\lfloor c/2 \rfloor + 1\bigl)] \text{ mod } N} \bra{x}
\end{equation}
that increments and decrements the position by the step size defined by the coin state. Because $C$ acts locally on the coin register, it is computationally inexpensive compared to the shift operator $S$. The shift operator performs conditional permutations across the entire position register, scaling with system size and thus constituting the primary resource bottleneck. For this reason, our work focuses exclusively on the efficient implementation of $S$.

\subsubsection{Increment/Decrement Scheme}
\begin{figure}[h!]
\centering
\begin{subfigure}{0.45\textwidth}
\centering
\begin{quantikz}[row sep=0.5cm, column sep=0.5cm]
    \lstick{$\ket{q_0^p}$} & \ctrl{1}\gategroup[5,steps=5,style={draw=none},label style={label position=above, yshift=0cm}]{{Increment}} & \ctrl{1} & \qw & \ \ldots\ & \ctrl{1} & \targ{} & \qw\\
    \lstick{$\ket{q_1^p}$} & \ctrl{2} & \ctrl{2} & \qw & \ \ldots\ & \targ{}  & \qw & \qw \\
    \lstick{\vdots} & \wireoverride{} & \wireoverride{} & \wireoverride{} & \wireoverride{} \reflectbox{$\ddots$} & \wireoverride{} & \wireoverride{} \\
    \lstick{$\ket{q_{n-2}^p}$} & \ctrl{1} & \targ{} & \qw & \ \ldots\ & \qw & \qw & \qw \\
    \lstick{$\ket{q_{n-1}^p}$} & \targ{} & \qw & \qw & \ \ldots\ & \qw & \qw & \qw
\end{quantikz}
\caption{}
\label{fig:increment_qc}
\end{subfigure}
\hfill
\begin{subfigure}{0.45\textwidth}
\centering
\begin{quantikz}[row sep=0.5cm, column sep=0.5cm]
    \lstick{$\ket{q_0^p}$} & \octrl{1}\gategroup[5,steps=5,style={draw=none},label style={label position=above, yshift=0cm}]{{Decrement}} & \octrl{1} & \qw & \ \ldots\ & \octrl{1} & \targ{} & \qw\\
    \lstick{$\ket{q_1^p}$} & \octrl{2} & \octrl{2} & \qw & \ \ldots\ & \targ{} & \qw & \qw \\
    \lstick{\vdots} & \wireoverride{} & \wireoverride{} & \wireoverride{} & \wireoverride{} \reflectbox{$\ddots$} & \wireoverride{} & \wireoverride{} \\
    \lstick{$\ket{q_{n-2}^p}$} & \octrl{1} & \targ{} & \qw & \ \ldots\ & \qw & \qw & \qw \\
    \lstick{$\ket{q_{n-1}^p}$} & \targ{} & \qw & \qw & \ \ldots\ & \qw & \qw & \qw
\end{quantikz}
\caption{}
\label{fig:decrement_qc}
\end{subfigure}
\hfill
\begin{subfigure}{0.45\textwidth}
\centering
\begin{quantikz}[row sep=0.5cm, column sep=0.5cm]
    \lstick{$\ket{q_0^p}$} & \qw & \gate[5]{Incr.}\gategroup[5,steps=2,style={dashed,rounded corners},label style={yshift=0.2cm}]{{\textbf{Shift Operator}}} & \gate[5]{Decr.} & \qw\\
    \lstick{$\ket{q_1^p}$} & \qw & & & \qw\\
    \vdots \\
    \lstick{$\ket{q_{n-2}^p}$} & \qw & & &\qw\\
    \lstick{$\ket{q_{n-1}^p}$} & \qw & & &\qw\\
    \lstick{$\ket{q_{0}^c}$} & \gate{C}\gategroup[1,steps=1,style={draw=none},label style={label position=below, yshift=-0.4cm}]{{\textbf{Coin Operator}}} & \octrl{-1} & \ctrl{-1} & \qw
\end{quantikz}
\caption{}
\label{fig:single-step_ID}
\end{subfigure}
\caption{Quantum circuit implementation of the Increment/Decrement scheme introduced in Douglas et al.~\cite{douglas_efficient_2009} (a) Increment operator acting on $n$ position qubits, implemented using multi-qubit controlled-NOT gates. (b) Decrement operator constructed analogously. (c) Single-step discrete-time quantum walk (DTQW) circuit, in which the increment and decrement subcircuits are conditionally applied based on the coin state $\ket{0}$ and $\ket{1}$, respectively.}
\label{fig:ID_scheme}
\end{figure}
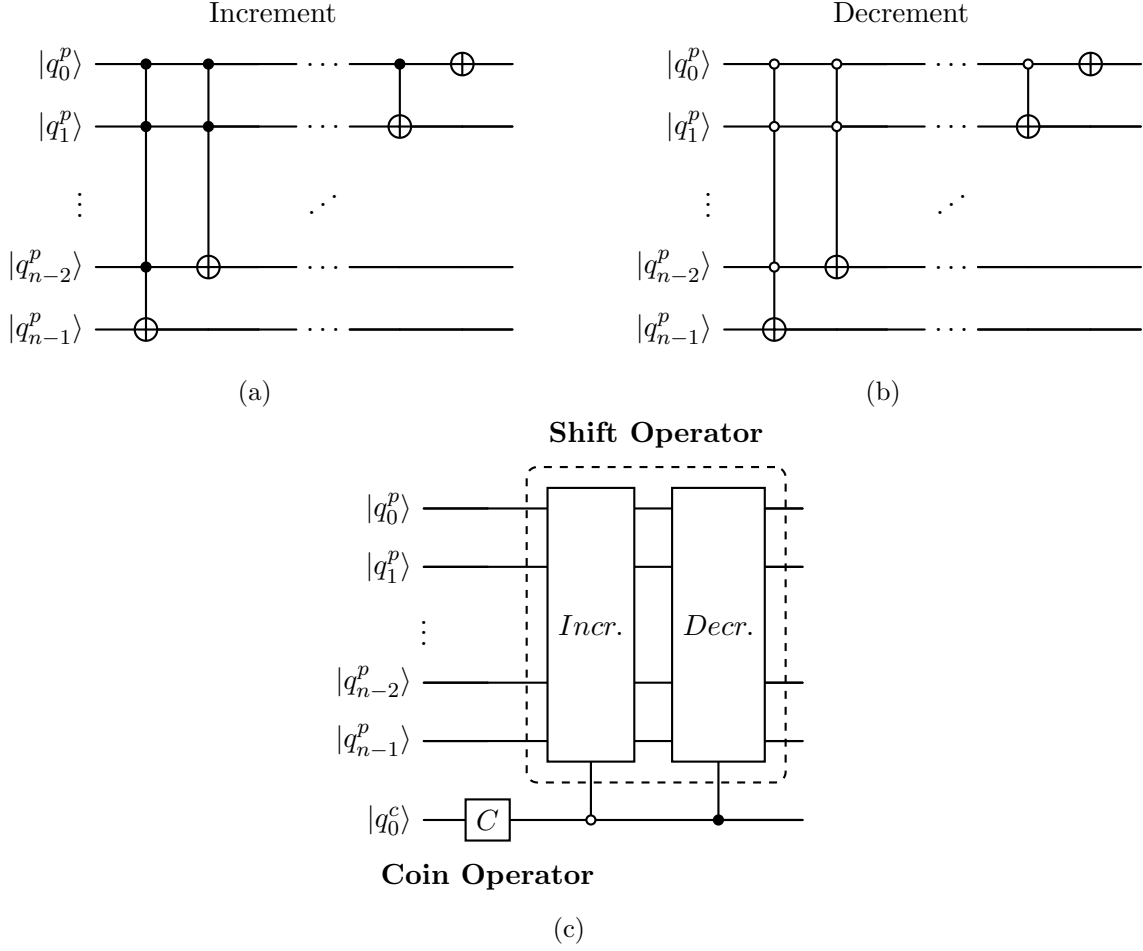

For an undirected $N$-cycle graph $\Gamma(\mathbb{Z}_N, \{1, -1\})$, the Hilbert space is simplified to $\mathcal{H}_{\text{cycle}} = \text{span}\bigr(\{\ket{c}\ket{x}:c=0,1; x=0,1,\dots, N-1\}\bigl)$. The coin states $\ket{0}$ and $\ket{1}$ correspond to incrementing ($\sigma=1$) and decrementing ($\sigma=-1$) the walker's position, respectively. The shift operator for this walk is defined as:
\begin{equation}
S = \ket{0}\bra{0}\otimes P_0 
   + \ket{1}\bra{1}\otimes P_1,
   \label{eq:dirac-shift-operator}
\end{equation}
where $P_0$ and $P_1$ are increment and decrement operators, notations borrowed from Razzoli et al.~\cite{razzoli_efficient_2024}. In a matrix representation, $S$ is a $2N\times2N$ block diagonal matrix:
\begin{equation}
    S = \begin{pmatrix}
    P_0 & 0 \\
    0 & P_1
    \end{pmatrix} = \begin{pmatrix}
    P_0 & 0 \\
    0 & P_0^{\mathsf{T}}
    \end{pmatrix},
    \label{eq:4}
\end{equation}
where $P_0$ and $P_1$ are $N\times N$ circulant matrices representing clockwise and anticlockwise shifts:
\begin{equation}
    P_0 = \begin{pmatrix}
        0 & 0 & 0 & \cdots & 1\\
        1 & \ddots & \ddots & \ddots & \vdots \\
        0 & \ddots & \ddots & 0 & 0 \\
        \vdots & \ddots & 1& 0 & 0 \\
        0  & \cdots & 0 & 1 & 0
    \end{pmatrix}, \qquad 
    P_1 = \begin{pmatrix}
        0 & 1 & 0 & \cdots & 0\\
        0 & \ddots & \ddots & \ddots & \vdots \\
        0 & \ddots & \ddots & 1 & 0 \\
        \vdots & \ddots & 0& 0 & 1 \\
        1  & \cdots & 0 & 0 & 0
    \end{pmatrix} = P_0^{\mathsf{T}}. 
    \label{eq:7}
\end{equation}
Douglas et al.~\cite{douglas_efficient_2009} directly maps these operators to a quantum circuit using multi-qubit controlled-NOT gates (Fig.~\ref{fig:increment_qc}, \ref{fig:decrement_qc}). The full circuit for a single time step of the DTQW is shown in Fig.~\ref{fig:single-step_ID}. However, this implementation model relies heavily on multi-qubit controlled operations, leading to large circuit depths that are costly to realize on NISQ devices. 

\subsubsection{Stepwise QFT Scheme}

Shakeel~\cite{shakeel_efficient_2020} improves upon the Increment/Decrement Scheme by decomposing the decrement operator as $P_1=JP_0J$, where $J$ is an $N\times N$ exchange matrix (anti-diagonal identity). Under this decomposition, the shift operator in Eq.~\ref{eq:4} becomes:
\begin{equation}
    S = \begin{pmatrix}
        {I}_N & 0\\
        0 & J
    \end{pmatrix}\begin{pmatrix}
        P_0 & 0\\
        0 & P_0
    \end{pmatrix}
    \begin{pmatrix}
        I_N & 0\\
        0 & J
    \end{pmatrix},
    \label{eq:6}
\end{equation}

\begin{figure}[h!]
\centering
\begin{tikzpicture}
    \node (circ) {
    \begin{quantikz}[row sep=0.45cm, column sep=0.44cm]
    \lstick{$\ket{q_0^p}$} & \qw & \qw\gategroup[6,steps=13,style={dashed,rounded corners},label style={yshift=0.1cm}]{{\textbf{Shift Operator}}} & \qw & \ \ldots\ & \qw & \targ{} & \gate[wires=5]{Incr.} & \targ{} & \qw & \qw & \ \ldots\ & \qw & \qw & \qw \\
    \lstick{$\ket{q_1^p}$} & \qw & \qw & \qw & \ \ldots\ & \targ{} & \qw & \ghost{Incr.} & \qw & \targ{} & \qw & \ \ldots\ & \qw & \qw & \qw \\
    \lstick{\vdots} & \wireoverride{} & \wireoverride{} & \wireoverride{} & \wireoverride{} \reflectbox{$\ddots$}\ & \wireoverride{} & \wireoverride{} & \wireoverride{} & \wireoverride{} & \wireoverride{} & \wireoverride{} & \wireoverride{} \ddots\ & \wireoverride{} & \wireoverride{} & \wireoverride{} \\  
    \lstick{$\ket{q_{n-2}^p}$} & \qw & \qw & \targ{} & \ \ldots\ & \qw & \qw & \ghost{Incr.} & \qw & \qw & \qw & \ \ldots\ & \targ{} & \qw & \qw \\
    \lstick{$\ket{q_{n-1}^p}$} & \qw & \targ{} & \qw & \ \ldots\ & \qw & \qw & \ghost{Incr.} & \qw & \qw & \qw & \ \ldots\ & \qw & \targ{} & \qw \\
    \lstick{$\ket{q_{0}^c}$} & \gate{C}\gategroup[1,steps=1,style={draw=none},label style={label position=below, yshift=-0.5cm}]{{\textbf{Coin Operator}}} & \ctrl{-1} & \ctrl{-2} & \ \ldots\ & \ctrl{-4} & \ctrl{-5} & \qw & \ctrl{-5} & \ctrl{-4} & \qw & \ \ldots\ & \ctrl{-2} & \ctrl{-1} & \qw 
    \end{quantikz}
    };
\end{tikzpicture}
\caption{Quantum circuit implementation of a single-step discrete-time quantum walk (DTQW) using the decomposed shift operator. The shift operator is realized through a sequence of controlled-NOT operations acting on the position register, with the control provided by the coin register, and an increment operator independent of the coin state. The coin operator is applied prior to the conditional shift, completing one time step of the walk.}
\label{fig:Improved-ID}
\end{figure}
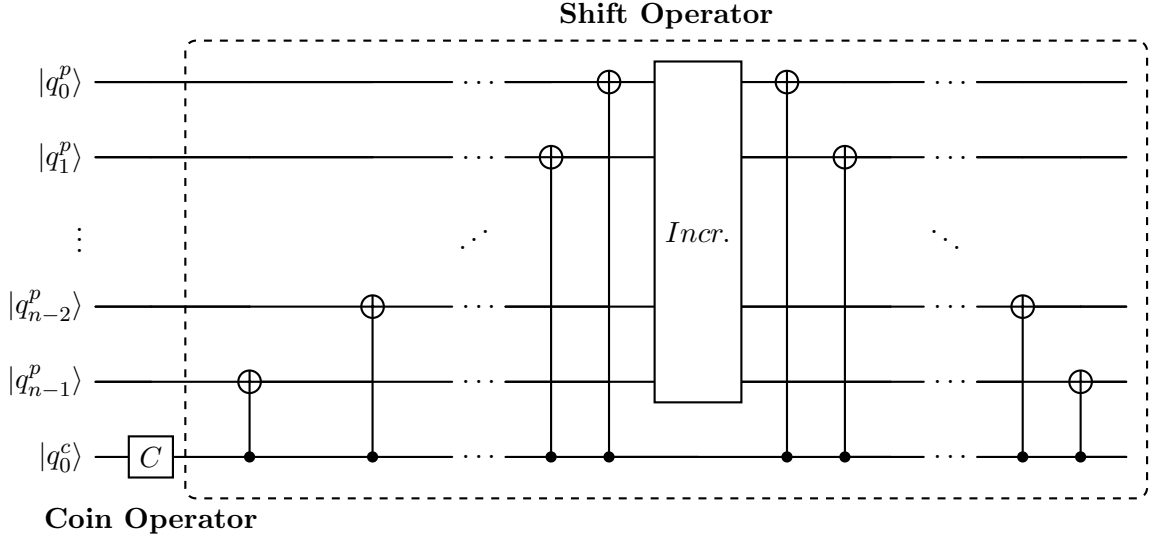

The outer matrices are implemented as a series of controlled-NOT gates, conditioned on the coin state $\ket{1}$. The central matrix corresponds to the increment operator in Fig.~\ref{fig:increment_qc} and is implemented independent of the coin state. The modified single time step circuit implementation is shown in Fig.~\ref{fig:Improved-ID}. 

To further optimize the construction, Shakeel diagonalizes the increment operator $P_0$ using the quantum Fourier Transform (QFT). As a circuit matrix, $P_0$ is diagonalized by the $N$-dimensional QFT matrix, defined as:
\begin{equation}
    \mathcal{F}_N = \frac{1}{\sqrt{N}} \begin{pmatrix}
        1 & 1 & 1 & 1 & \cdots & 1 \\
        1 & \omega & \omega^2 & \omega^3 & \cdots & \omega^{N-1}\\
        1 & \omega^2 & \omega^4 & \omega^6 & \cdots & \omega^{2(N-1)} \\
        1 & \omega^3 & \omega^6 & \omega^9 & \cdots & \omega^{3(N-1)} \\
        \vdots & \vdots & \vdots & \vdots & \ddots & \vdots \\
        1 & \omega^{N-1} & \omega^{2(N-1)} & \omega^{3(N-1)} & \cdots & \omega^{(N-1)(N-1)}
    \end{pmatrix}.
    \label{eq:12}
\end{equation}
where $\omega_N = e^{2\pi i/N}$. Consequently, the operators and rewritten as:
\begin{equation}
    P_0 = \mathcal{F}^\dagger \Omega^\dagger\mathcal{F}, \quad \text{and} \quad P_1 =\mathcal{F}^\dagger \Omega \mathcal{F},
\end{equation}
where $\Omega$ is a diagonal phase matrix:
\begin{equation}
    \Omega = \text{diag}\bigl(1, \omega_N, \omega^2_N, \dots, \omega^{N-1}_N\bigr).
\end{equation}
Implementing $\Omega$ requires $n$ rotation gates are required, given by:
\begin{align}
\Omega &= \bigotimes_{\ell=0}^{n-1} R_{\ell+1} =R_1 \otimes R_2 \otimes \cdots \otimes R_n, 
\end{align}
where $\ell$ is the index of the position qubit, and the  rotation gate $R_\lambda$ is defined as:
\begin{equation}
    R_\lambda = \begin{pmatrix}
        1 & 0 \\
        0 & \omega^{2^{n-\lambda}}_N
    \end{pmatrix} = \begin{pmatrix}
        1 & 0 \\
        0 & e^{\frac{2\pi i}{2^\lambda}}
    \end{pmatrix}.
\end{equation}

Further efficiency is achieved by eliminating the SWAP operations $\uptau$ from the QFT implementation.
\begin{equation}
    \uptau : \ket{q_{n-1}^p,q_{n-2}^p,\dots, q_0^p} \mapsto \ket{q_0^p, q_1^p, \dots q_{n-1}^p}.
\end{equation}
Let $ \tilde{\mathcal{F}} = \uptau \mathcal{F}$ and $\tilde{\mathcal{F}}^\dagger = \mathcal{F}^\dagger \uptau$ denote SWAP-free QFT and Inverse QFT (IQFT). The diagonal phase matrix in this basis becomes:
\begin{equation}
    \tilde{\Omega} = \uptau \Omega \uptau = \bigotimes_{\ell=n-1}^0 R_{\ell+1} = R_n \otimes R_{n-1} \otimes \dots \otimes R_1.
    \label{eq:19}
\end{equation}
Applying this to the shift operator, Eq.~\ref{eq:6} becomes:
\begin{equation}
    S = \begin{pmatrix}
        {I}_N & 0\\
        0 & J
    \end{pmatrix}\begin{pmatrix}
        \tilde{\mathcal{F}}^\dagger & 0\\
        0 & \tilde{\mathcal{F}}^\dagger
    \end{pmatrix}
    \begin{pmatrix}
        \tilde{\Omega}^\dagger & 0\\
        0 & \tilde{\Omega}^\dagger
    \end{pmatrix}\begin{pmatrix}
        \tilde{\mathcal{F}} & 0\\
        0 & \tilde{\mathcal{F}}
    \end{pmatrix}
    \begin{pmatrix}
        I_N & 0\\
        0 & J
    \end{pmatrix},
\end{equation}

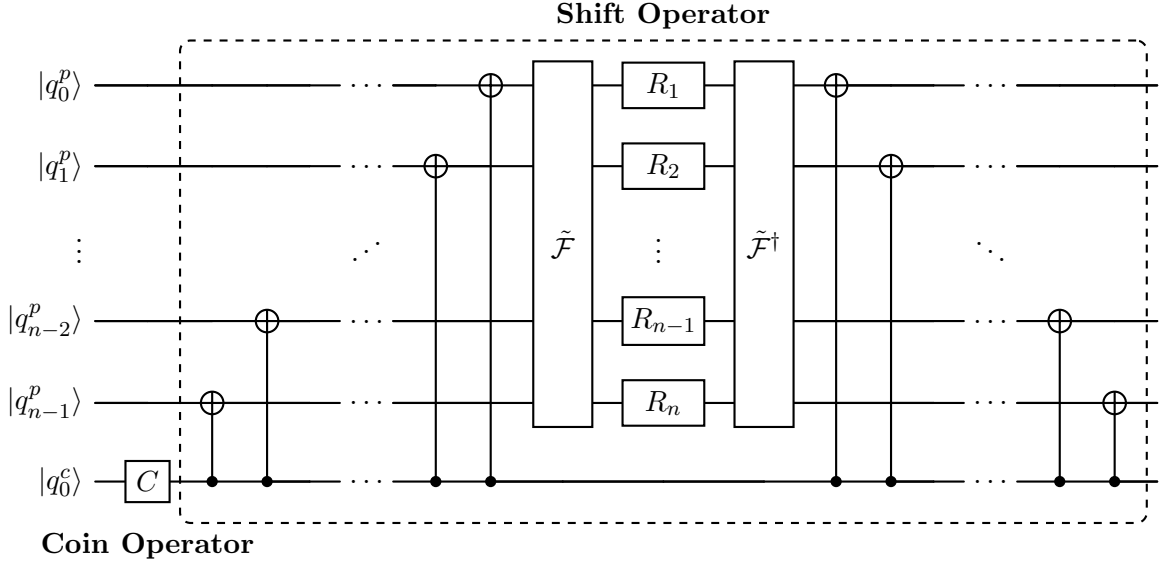
\begin{figure}[h]
\centering
\begin{tikzpicture}
    \node (circ) {
    \tikzset{Rgate/.style={draw, minimum width=0.7cm}}
    \begin{quantikz}[row sep=0.45cm, column sep=0.4cm]
    \lstick{$\ket{q_0^p}$} & \qw & \qw\gategroup[6,steps=15,style={dashed,rounded corners},label style={yshift=0.1cm}]{{\textbf{Shift Operator}}} & \qw & \qw &  \ \ldots\ & \qw & \targ{} & \gate[wires=5]{\makebox[0.5cm]{$\tilde{\mathcal{F}}$}} & \gate{\makebox[0.8cm]{\(R_1\)}} & \gate[wires=5]{\makebox[0.5cm]{$\tilde{\mathcal{F}}^\dagger$}} & \targ{} & \qw & \qw &  \ \ldots\ & \qw & \qw & \qw \\
    \lstick{$\ket{q_1^p}$} & \qw & \qw & \qw & \qw &  \ \ldots\ & \targ{} & \qw & & \gate{\makebox[0.8cm]{\(R_2\)}} & & \qw & \targ{} & \qw &  \ \ldots\ & \qw & \qw & \qw \\
    \lstick{\vdots} & \wireoverride{} & \wireoverride{} & \wireoverride{} & \wireoverride{} & \wireoverride{}\ \reflectbox{$\ddots$}\ & \wireoverride{} & \wireoverride{} & \wireoverride{} & \wireoverride{} \vdots\ & \wireoverride{} & \wireoverride{} & \wireoverride{} & \wireoverride{} & \wireoverride{}\ \ddots\ & \wireoverride{} & \wireoverride{} & \wireoverride{} \\  
    \lstick{$\ket{q_{n-2}^p}$} & \qw & \qw & \targ{} & \qw &  \ \ldots\ & \qw & \qw & & \gate{\makebox[0.8cm]{\(R_{n-1}\)}} & & \qw & \qw & \qw &  \ \ldots\ & \targ{} & \qw & \qw \\
    \lstick{$\ket{q_{n-1}^p}$} & \qw & \targ{} & \qw & \qw &  \ \ldots\ & \qw & \qw & & \gate{\makebox[0.8cm]{\(R_n\)}} & & \qw & \qw & \qw &  \ \ldots\ & \qw & \targ{} & \qw \\
    \lstick{$\ket{q_{0}^c}$} & \gate{C}\gategroup[1,steps=1,style={draw=none},label style={label position=below, yshift=-0.5cm}]{{\textbf{Coin Operator}}} & \ctrl{-1} & \ctrl{-2} & \qw &  \ \ldots\ & \ctrl{-4} & \ctrl{-5} & \qw & \qw & \qw & \ctrl{-5} & \ctrl{-4} & \qw & \ \ldots\ & \ctrl{-2} & \ctrl{-1} & \qw 
    \end{quantikz}
    };
\end{tikzpicture}
\caption{Final quantum circuit implementation of a single-step DTQW in the stepwise QFT scheme~\cite{shakeel_efficient_2020}, showing the decomposition of the increment operator using QFT, phase rotations $R_\lambda$, and the IQFT.}
\label{fig:stepwise_QFT}
\end{figure}

The resulting single time step circuit is shown in Fig.~\ref{fig:stepwise_QFT}. While this scheme reduces the complexity of the increment operator, it requirest the implementation of the QFT and IQFT at every time step, which must be repeated $t$ times for the full evolution.

\subsubsection{Boundary QFT Scheme}
The current state-of-the-art circuit implementation model, proposed by Razzoli et al.~\cite{razzoli_efficient_2024}, builds upon the previous schemes by directly diagonalizing the $2N\times2N$ shift operator given in Eq.~\ref{eq:4}. By applying the SWAP-free QFT, the shift operator yields a diagonal phase shift operator $\Sigma$:
\begin{equation}
    \Sigma = \begin{pmatrix}
        \tilde{\mathcal{F}} & 0 \\
        0 & \tilde{\mathcal{F}}
    \end{pmatrix}\begin{pmatrix}
        P_0 & 0\\
        0 & P_1
    \end{pmatrix}\begin{pmatrix}
        \tilde{\mathcal{F}}^\dagger & 0 \\
        0 & \tilde{\mathcal{F}}^\dagger
    \end{pmatrix} = \begin{pmatrix}
        \tilde{\Omega}^\dagger & 0\\
        0 & \tilde{\Omega}
    \end{pmatrix}.
\end{equation}
Consequently, the shift operator $S$ in Eq.~\ref{eq:dirac-shift-operator} can be reformulated as:
\begin{equation}
    S = \left(I_k\otimes \tilde{\mathcal{F}}^\dagger\right)\left( \ket{0}\bra{0}\otimes \tilde{\Omega}^\dagger + \ket{1}\bra{1}\otimes \tilde{\Omega}\right)\left(I_k\otimes \tilde{\mathcal{F}}\right),
\end{equation}
where $I_k$ is the identity operator on the coin space. This requires $2n$ controlled-rotation gates ($CR_{\ell+1}$) to implement $\tilde{\Omega}^\dagger$ and $\tilde{\Omega}$. To optimize this, Razzoli et al. decompose $\Sigma$ to reduce the count of multi-qubit gates: 
\begin{align}
    \Sigma &= \begin{pmatrix}
        I_N & 0 \\
        0 & \tilde{\Omega}^2
    \end{pmatrix}\begin{pmatrix}
        \tilde{\Omega}^\dagger & 0 \\
        0 & \tilde{\Omega}^\dagger
    \end{pmatrix} \notag \\
    &= \left(\ket{0}\bra{0}\otimes I_N + \ket{1}\bra{1}\otimes \tilde{\Omega}^2\right)\left(I_k\otimes \tilde{\Omega}^\dagger\right).
    \label{eq:decomposed-phase-shift-operator}
\end{align}
Since $R_\lambda^2 = R_{\lambda-1}$ and $R_0=I$, the term $\tilde{\Omega}^2$ simplifies to:
\begin{equation}
    \tilde{\Omega}^2 = \bigotimes_{\ell=n-1}^0 R_{\ell+1}^2 = \bigotimes_{\ell=n-1}^0 R_{\ell}= R_{n-1} \otimes \dots \otimes R_1 \otimes I, 
    \label{eq:24}
\end{equation}
only requiring $n-1$ rotation gates to implement.

This decomposed operator in Eq.~\ref{eq:decomposed-phase-shift-operator} requires only $n-1$ $CR_\ell$ gates for $\tilde{\Omega}^2$, while the $\tilde{\Omega}\dagger$ term is implemented using $n$ single-qubit rotations, as it is now independent of the coin state. 

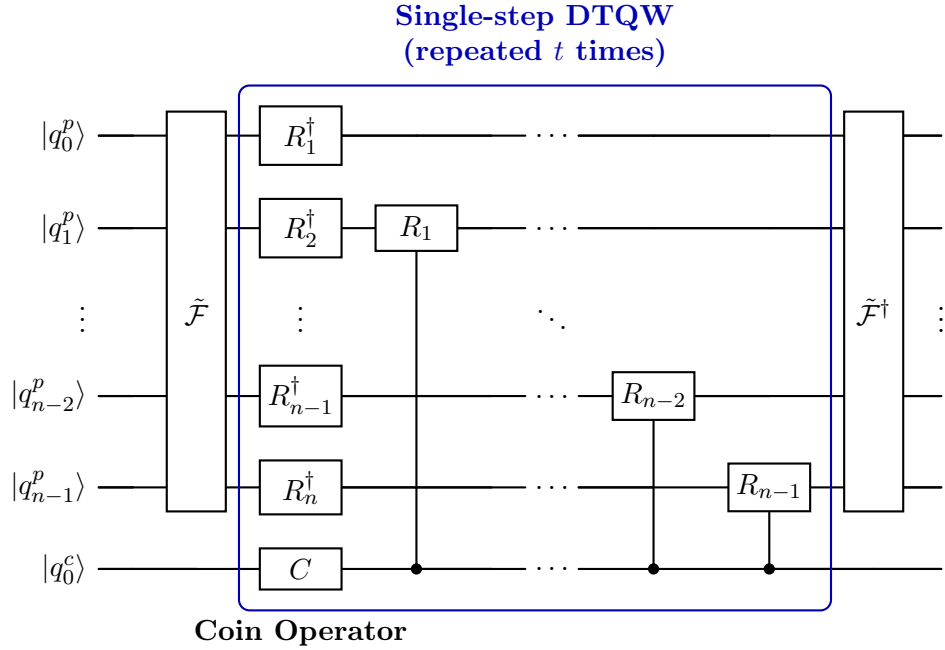
\begin{figure}[h]
\centering
\begin{tikzpicture}
    \node (circ) {
    \tikzset{Rgate/.style={draw, minimum width=0.7cm}}
    \begin{quantikz}[row sep=0.45cm, column sep=0.45cm]
    \lstick{$\ket{q_0^p}$} & \qw & \gate[wires=5]{\makebox[0.5cm]{$\tilde{\mathcal{F}}$}} & \gate{\makebox[0.8cm]{\(R_1^\dagger\)}}\gategroup[6,steps=6,style={blue!70!black, rounded corners},label style={yshift=0.2cm, blue!70!black}]{\shortstack{\textbf{Single-step DTQW}\\\textbf{(repeated $t$ times)}}} & \qw & \qw &  \ \ldots\ & \qw & \qw & \gate[wires=5]{\makebox[0.5cm]{$\tilde{\mathcal{F}}^\dagger$}} & \qw \\
    \lstick{$\ket{q_1^p}$} & \qw & & \gate{\makebox[0.8cm]{\(R_2^\dagger\)}} & \gate{\makebox[0.8cm]{\(R_1\)}} & \qw &  \ \ldots\ & \qw & \qw & \qw & \qw \\
    \lstick{\vdots} & \wireoverride{} & \wireoverride{} & \wireoverride{} \vdots & \wireoverride{} & \wireoverride{} & \wireoverride{}\ \ddots\ & \wireoverride{} & \wireoverride{} & \wireoverride{} & \wireoverride{} \vdots \\  
    \lstick{$\ket{q_{n-2}^p}$} & \qw & & \gate{\makebox[0.8cm]{\(R_{n-1}^\dagger\)}} & \qw & \qw &  \ \ldots\ & \gate{\makebox[0.8cm]{\(R_{n-2}\)}} & \qw & & \qw \\
    \lstick{$\ket{q_{n-1}^p}$} & \qw & & \gate{\makebox[0.8cm]{\(R_n^\dagger\)}} & \qw & \qw &  \ \ldots\ & \qw & \gate{\makebox[0.8cm]{\(R_{n-1}\)}} & & \qw \\
    \lstick{$\ket{q_{0}^c}$} & \qw & \qw & \gate{\makebox[0.8cm]{\(C\)}}\gategroup[1,steps=1,style={draw=none},label style={label position=below, yshift=-0.5cm}]{{\textbf{Coin Operator}}} & \ctrl{-4} & \qw &  \ \ldots\  &\ctrl{-2} & \ctrl{-1} & \qw & \qw
    \end{quantikz}
    };
\end{tikzpicture}
\caption{Quantum circuit implementation of DTQW in the boundary QFT scheme~\cite{razzoli_efficient_2024}. The QFT and its inverse are implemented only once at the start and the end excluding the SWAP operation. The central block corresponds to the single time step and is to be repeated $t$ times.}
\label{fig:boundary_QFT_qc}
\end{figure}

The defining advantage of this scheme is realized during time evolution. The total unitary evolution $U^t$ forms a telescoping product:
\begin{align}
    U^t &= \left[\left(I_k\otimes \tilde{\mathcal{F}}^\dagger\right)\left[\left( \ket{0}\bra{0}\otimes I_N + \ket{1}\bra{1}\otimes \tilde{\Omega}^2\right)\left(C\otimes \tilde{\Omega}^\dagger\right)\right]\left(I_k\otimes \tilde{\mathcal{F}}\right)\right]^t \notag \\
    &=\left(I_k\otimes \tilde{\mathcal{F}}^\dagger\right)\left[\left(\ket{0}\bra{0}\otimes I_N + \ket{1}\bra{1}\otimes \tilde{\Omega}^2\right)\left(C\otimes \tilde{\Omega}^\dagger\right)\right]^t\left(I_k\otimes \tilde{\mathcal{F}}\right)
\end{align}
Because the intermediate QFT/IQFT stages cancel out ($\tilde{\mathcal{F}}\tilde{\mathcal{F}}^\dagger=I$), the QFT and IQFT are implemented only once at the "boundaries"---at the very beginning and the end of the circuit---as shown in Fig.~\ref{fig:boundary_QFT_qc}. This significant gate reduction is unattainable in the Stepwise QFT scheme, as the interleaving of the coin operator and the controlled-NOT gates prevents the QFT and IQFT from canceling between successive time steps.

\section{DTQW on the N-Cayley graph}
\label{1D_cayley_graph}
We now generalize the DTQW framework, specifically the Boundary QFT Scheme by Razzoli et al, to a 1D Cayley graph $\Gamma(\mathbb{Z}_n, \mathcal{S)}$, where the connectivity is determined by a generating set $\mathcal{S}=\{\sigma_0, \sigma_1, \dots,\sigma_{k-1}\}$. The construction of the shift operator on such graphs depends fundamentally on the algebraic structure of $\mathcal{S}$. Specifically, the construction can fall into two primary categories: inverse-closed generating sets, which correspond to undirected graphs, and non-inverse-closed sets, which describe directed graphs.

Within the inverse-closed regime, the implementation further depends on whether the generating set involves an involution (a self-inverse element where $\sigma \equiv -\sigma \pmod N$). While the non-involution case allows for a simpler mapping to quantum logic due to the grouping of inverse pairs ($\sigma,-\sigma$), the methodology can be extended to involutory sets with an additional padding. To illustrate the reduction in gate complexity, we provide a detailed implementation for an 8-Cayley graph with an involutory generating set.

\subsection{Inverse-closed $\mathcal{S}$ without an involution}
We first consider the case where the generating set $\mathcal{S}$ is inverse-closed and contains no involution ($\sigma \not\equiv -\sigma \pmod N$ for all $\sigma \in \mathcal{S}$). $\mathcal{S}$ must consist of consecutive integer increments and their respective inverses. Formally, we define the generating set as: $\mathcal{S}=\{1, -1, 2, -2, \dots, k/2, -k/2\}$, where $k=|\mathcal{S}|$ is the degree of the graph, assumed to be a power of 2 for circuit compatibility.

The shift operator for the resulting Cayley graph $\Gamma\left( \mathbb{Z}_N, \mathcal{S} \right)$ is represented as a $kN \times kN$ block diagonal matrix: 
\begin{equation}
    S = \begin{pmatrix}
     P_0 & & & & & & & & \\
     & P_1 & & & & & & & \\
     & & P_2 & & & & & & \\
     & & & P_3 & & & & & \\
     & & & & P_4 & & & & \\
     & & & & & P_5 & & & \\
     & & & & & & \ddots & & \\
     & & & & & & & P_{k-2} & \\
     & & & & & & & & P_{k-1}
    \end{pmatrix},
    \label{eq:26}
\end{equation}
where each $P_c$ corresponds to the increment / decrement (ID) operator associated with coin state $\ket{c}$. Specifically, $P_c$ shifts the walker's position by a step size $\sigma$ defined by the generators. For instance, the operators for $\sigma=2,-2$ (coin states $c=2,3$) are given by:
\begin{equation}
    P_2 = \begin{pmatrix}
        0 & \cdots & 0 & 1 & 0\\
        0 & \ddots & \ddots & \ddots & 1 \\
        1 & \ddots & \ddots & 0 & 0 \\
        \vdots & \ddots & 0 & 0 & \vdots \\
        0  & \cdots & 1 & 0 & 0
    \end{pmatrix}, \qquad 
    P_3 = \begin{pmatrix}
        0 & 0 & 1 & \cdots & 0\\
        \vdots & \ddots & \ddots & \ddots & \vdots \\
        0 & \ddots & \ddots & 0 & 1 \\
        1 & \ddots & 0& 0 & 0 \\
        0  & 1 & 0 & \cdots & 0
    \end{pmatrix} = P_2^{\mathsf{T}}. 
    \label{eq:27}
\end{equation}

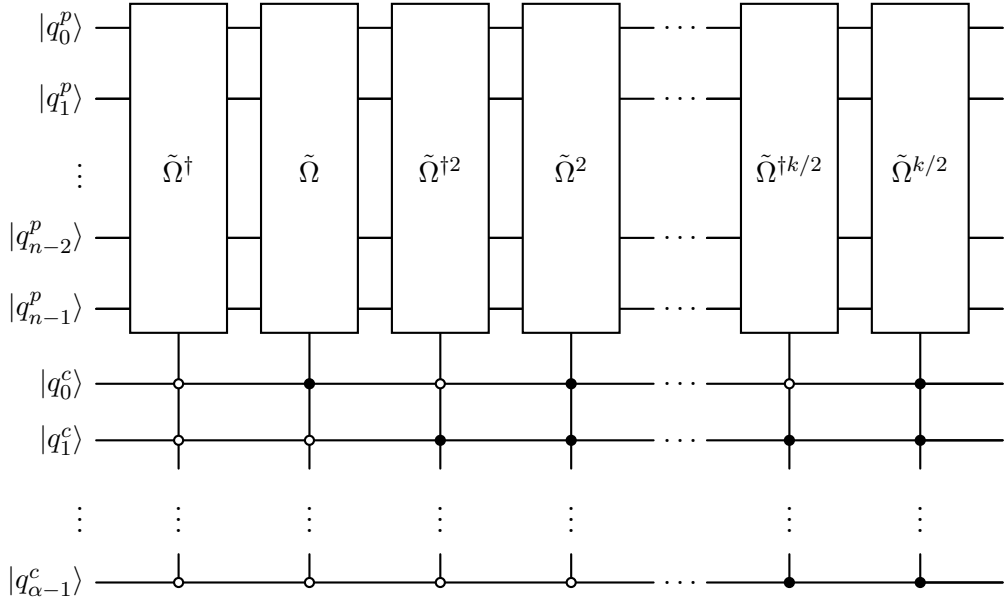
\begin{figure}
\centering
\begin{tikzpicture}
    \node (circ) {
    \tikzset{Rgate/.style={draw, minimum width=0.7cm}}
    \begin{quantikz}[row sep=0.6cm, column sep=0.45cm]
    \lstick{$\ket{q_0^p}$}  & \gate[wires=5]{\makebox[1cm]{$\tilde{\Omega}^\dagger$}} & \gate[wires=5]{\makebox[1cm]{$\tilde{\Omega}$}} & \gate[wires=5]{\makebox[1cm]{$\tilde{\Omega}^{\dagger2}$}} & \gate[wires=5]{\makebox[1cm]{$\tilde{\Omega}^{2}$}} & \ \ldots\ & \gate[wires=5]{\makebox[1cm]{$\tilde{\Omega}^{\dagger k/2}$}} & \gate[wires=5]{\makebox[1cm]{$\tilde{\Omega}^{k/2}$}} & \qw \\
    \lstick{$\ket{q_1^p}$} & \qw & \qw & \qw & \qw & \ \ldots\ & \qw & \qw & \qw \\ [0.3cm]
    \lstick{\vdots} & \wireoverride{} & \wireoverride{} & \wireoverride{} & \wireoverride{} & \wireoverride{} & \wireoverride{} & \wireoverride{} & \wireoverride{} \\ [0.3cm] 
    \lstick{$\ket{q_{n-2}^p}$} & \qw & \qw & \qw & \qw & \ \ldots\ & \qw & \qw & \qw \\
    \lstick{$\ket{q_{n-1}^p}$} & \qw & \qw & \qw & \qw & \ \ldots\ & \qw & \qw & \qw \\
    \lstick{$\ket{q_{0}^c}$} & \octrl{-1} & \ctrl{-1} & \octrl{-1} & \ctrl{-1} & \ \ldots\ & \octrl{-1} & \ctrl{-1} & \qw \\
    \lstick{$\ket{q_{1}^c}$} & \octrl{-1}\wire[d]{q} & \octrl{-1}\wire[d]{q} & \ctrl{-1}\wire[d]{q} & \ctrl{-1}\wire[d]{q} & \ \ldots\ & \ctrl{-1}\wire[d]{q} & \ctrl{-1}\wire[d]{q} & \qw \\ [-0.3cm]
    \wireoverride{} & \wireoverride{} & \wireoverride{} & \wireoverride{} & \wireoverride{} & \wireoverride{} & \wireoverride{} & \wireoverride{} & \wireoverride{} \\ [-0.3cm]
    \lstick{\vdots} & \wireoverride{} \vdots & \wireoverride{} \vdots & \wireoverride{} \vdots & \wireoverride{} \vdots & \wireoverride{} & \wireoverride{} \vdots & \wireoverride{} \vdots & \wireoverride{} \\ [-0.3cm]
    \wireoverride{} & \wireoverride{} & \wireoverride{} & \wireoverride{} & \wireoverride{} & \wireoverride{} & \wireoverride{} & \wireoverride{} & \wireoverride{} \\ [-0.3cm]
    \lstick{$\ket{q_{\alpha-1}^c}$} & \octrl{-1} & \octrl{-1} & \octrl{-1} & \octrl{-1} & \ \ldots\ & \ctrl{-1} & \ctrl{-1} & \qw
    \end{quantikz}
    };
\end{tikzpicture}
\caption{Straightforward quantum circuit implementation of the phase shift operator $\Sigma$ in Eq.~\ref{eq:30}.}
\label{fig:naive_qc}
\end{figure}

Since the ID operators $P_c$ are circulant matrices, each odd-indexed operator is the transpose of its even-index predecessor ($P_c = P_{c-1}^\mathsf{T}$ for every odd $c$), and all ID operators can be expressed as powers of the fundamental unit increment operator $P_0$:
\begin{equation}
    P_c = 
    \begin{cases}
    P_0^{c/2+1} \qquad \text{for even c}\\
    P_0^{\mathsf{T} (c+1)/2} \qquad \text{for odd c}
    \end{cases}
\end{equation}

Under this formulation, the shift operator from Eq.~\ref{eq:26} can be rewritten as:
\begin{equation}
    S = \text{diag} \left( P_0, P_0^{\mathsf{T}},P_0^2, P_0^{\mathsf{T}2}, P_0^3, P_0^{\mathsf{T}3}, \dots , P_0^{(k/2)}, P_0^{\mathsf{T}(k/2)} \right).
\end{equation}

Applying the SWAP-free QFT and IQFT matrices diagonalize the shift operator: $S = (I_k\otimes \tilde{\mathcal{F}}^\dagger)\Sigma(I_k\otimes \tilde{\mathcal{F}})$. The resulting phase shift operator $\Sigma$ is a $kN\times kN$ block operator:
\begin{equation}
    \Sigma = \text{diag}\left(\tilde{\Omega}^\dagger, \tilde{\Omega}, \tilde{\Omega}^{\dagger2}, \tilde{\Omega}^2,\tilde{\Omega}^{\dagger3}, \tilde{\Omega}^3,\dots, \tilde{\Omega}^{\dagger (k/2)}, \tilde{\Omega}^{(k/2)} \right),
    \label{eq:30}
\end{equation}
where $\tilde{\Omega}$ is the SWAP-free diagonal phase matrix as defined in Eq.~\ref{eq:19}. A direct implementation of $\Sigma$, shown in Fig.~\ref{fig:naive_qc}, requires $O(kn)$ multi-controlled rotation gates ($C^{(\alpha)}R_\lambda$), where each gate is conditioned on the full $\alpha$-qubit coin register to implement specific rotations. To improve the complexity of multi-controlled gates, we decompose $\Sigma$ in Eq.~\ref{eq:30} hierarchically. 

\subsubsection*{First-order decomposition}
In the first-order decomposition, we sequentially factor out $2n \times2n$ block diagonal matrices composed of $\tilde{\Omega}^\dagger$ and $\tilde{\Omega}$. Specifically, we express $\Sigma$ as a product of $k/2$ block diagonal matrices $M_j$, where each matrix $M_j$ is structured such that the first $2j$ diagonal blocks are identity matrices, while the remaining blocks consist of alternating pairs of $\tilde{\Omega}^\dagger$ and $\tilde{\Omega}$: 
\begin{equation}
    \Sigma = \prod_{j=0}^{k/2-1}M_j, \quad \text{where} \quad M_j=\text{diag}\left(\underbrace{I, \dots, I}_{\text{$2j$}},\underbrace{(\tilde{\Omega}^\dagger, \tilde{\Omega}),\dots, (\tilde{\Omega}^\dagger, \tilde{\Omega})}_{\text{$k/2-j$}}\right)
    \label{eq:31}
\end{equation}
Explicitly, the decomposition proceeds as
\begin{align}
    \Sigma = \scalebox{0.55}{$\underbrace{\begin{pmatrix}
     \tilde{\Omega}^\dagger & & & & & & & & \\
     & \tilde{\Omega} & & & & & & & \\
     & & \tilde{\Omega}^\dagger & & & & & & \\
     & & & \tilde{\Omega} & & & & & \\
     & & & & \tilde{\Omega}^\dagger & & & & \\
     & & & & & \tilde{\Omega} & & & \\
     & & & & & & \ddots & & \\
     & & & & & & & \tilde{\Omega}^\dagger & \\
     & & & & & & & & \tilde{\Omega}
     \end{pmatrix}}_{\text{\LARGE $M_0$}}\underbrace{\begin{pmatrix}
     I & & & & & & & & \\
     & I & & & & & & & \\
     & & \tilde{\Omega}^\dagger & & & & & & \\
     & & & \tilde{\Omega} & & & & & \\
     & & & & \tilde{\Omega}^\dagger & & & & \\
     & & & & & \tilde{\Omega} & & & \\
     & & & & & & \ddots & & \\
     & & & & & & & \tilde{\Omega}^\dagger & \\
     & & & & & & & & \tilde{\Omega}
     \end{pmatrix}}_{\text{\LARGE$M_1$}}\underbrace{\begin{pmatrix}
     I & & & & & & & & \\
     & I & & & & & & & \\
     & & I & & & & & & \\
     & & & I & & & & & \\
     & & & & \tilde{\Omega}^\dagger & & & & \\
     & & & & & \tilde{\Omega} & & & \\
     & & & & & & \ddots & & \\
     & & & & & & & \tilde{\Omega}^\dagger & \\
     & & & & & & & & \tilde{\Omega}
     \end{pmatrix}}_{\text{\LARGE$M_2$}}\hdots\underbrace{\begin{pmatrix}
     I & & & & & & & & \\
     & I & & & & & & & \\
     & & I & & & & & & \\
     & & & I & & & & & \\
     & & & & I & & & & \\
     & & & & & I & & & \\
     & & & & & & \ddots & & \\
     & & & & & & & \tilde{\Omega}^\dagger & \\
     & & & & & & & & \tilde{\Omega}
\end{pmatrix}}_{\text{\LARGE$M_{k/2-1}$}}$}
\label{eq:32}
\end{align}

\subsubsection*{Second-order decomposition}
\label{Second-order decomposition}
We further decompose the $2n\times2n$ blocks of ($\tilde{\Omega}^\dagger, \tilde{\Omega}$) in $M_j$, following the same approach as in Eq.~\ref{eq:decomposed-phase-shift-operator}. This factorization allows us to redefine each matrix $M_j$ as a product of two fundamental matrix structures, $M_j = A_jD_j$:
\begin{equation}
    A_j = \text{diag}\left(\underbrace{I, \dots, I}_{\text{$2j$}},\underbrace{(I, \tilde{\Omega}^2),\dots, (I, \tilde{\Omega}^2)}_{\text{$k/2-j$}}\right),\quad D_j = \text{diag}\left(\underbrace{I, \dots, I}_{\text{$2j$}},\underbrace{(\tilde{\Omega}^\dagger, \tilde{\Omega}^\dagger),\dots, (\tilde{\Omega}^\dagger, \tilde{\Omega}^\dagger)}_{\text{$k/2-j$}}\right).
\end{equation}

The phase shift operator $\Sigma$ from Eq.~\ref{eq:31} is then rewritten as:
\begin{equation}
    \Sigma = \prod_{j=0}^{k/2-1} \left(A_jD_j\right), 
    \label{eq:33}
\end{equation}
and the explicit decomposition of $\Sigma$ in Eq.~\ref{eq:32} is shown as
\begin{equation}
\begin{aligned}
     \Sigma = & \scalebox{0.53}{$
     \underbrace{\begin{pmatrix}
     I & & & & & & & & \\
     & \tilde{\Omega}^2 & & & & & & & \\
     & & I & & & & & & \\
     & & & \tilde{\Omega}^2 & & & & & \\
     & & & & I & & & & \\
     & & & & & \tilde{\Omega}^2 & & & \\
     & & & & & & \ddots & & \\
     & & & & & & & I & \\
     & & & & & & & & \tilde{\Omega}^2
    \end{pmatrix}}_{\text{\LARGE$A_0$}}
    \underbrace{\begin{pmatrix}
    \begin{array}{@{\hskip 2.8pt}c@{\hskip 2.8pt} @{\hskip 2.8pt}c@{\hskip 2.8pt} @{\hskip 2.8pt}c@{\hskip 2.8pt} @{\hskip 2.8pt}c@{\hskip 2.8pt} @{\hskip 2.8pt}c@{\hskip 2.8pt} @{\hskip 2.8pt}c@{\hskip 2.8pt} @{\hskip 2.8pt}c@{\hskip 2.8pt} @{\hskip 2.8pt}c@{\hskip 2.8pt} @{\hskip 2.8pt}c@{\hskip 2.8pt}}
     \tilde{\Omega}^\dagger & & & & & & & & \\
     & \tilde{\Omega}^\dagger & & & & & & & \\
     & & \tilde{\Omega}^\dagger & & & & & & \\
     & & & \tilde{\Omega}^\dagger & & & & & \\
     & & & & \tilde{\Omega}^\dagger & & & & \\
     & & & & & \tilde{\Omega}^\dagger & & & \\
     & & & & & & \ddots & & \\
     & & & & & & & \tilde{\Omega}^\dagger & \\
     & & & & & & & & \tilde{\Omega}^\dagger
    \end{array}
    \end{pmatrix}}_{\text{\LARGE$D_0$}}
    \underbrace{\begin{pmatrix}
     I & & & & & & & & \\
     & I & & & & & & & \\
     & & I & & & & & & \\
     & & & \tilde{\Omega}^2 & & & & & \\
     & & & & I & & & & \\
     & & & & & \tilde{\Omega}^2 & & & \\
     & & & & & & \ddots & & \\
     & & & & & & & I & \\
     & & & & & & & & \tilde{\Omega}^2
    \end{pmatrix}}_{\text{\LARGE$A_1$}}
    \underbrace{\begin{pmatrix}
    \begin{array}{@{\hskip 2.8pt}c@{\hskip 2.8pt} @{\hskip 2.8pt}c@{\hskip 2.8pt} @{\hskip 2.8pt}c@{\hskip 2.8pt} @{\hskip 2.8pt}c@{\hskip 2.8pt} @{\hskip 2.8pt}c@{\hskip 2.8pt} @{\hskip 2.8pt}c@{\hskip 2.8pt} @{\hskip 2.8pt}c@{\hskip 2.8pt} @{\hskip 2.8pt}c@{\hskip 2.8pt} @{\hskip 2.8pt}c@{\hskip 2.8pt}}
     I & & & & & & & & \\
     & I & & & & & & & \\
     & & \tilde{\Omega}^\dagger & & & & & & \\
     & & & \tilde{\Omega}^\dagger & & & & & \\
     & & & & \tilde{\Omega}^\dagger & & & & \\
     & & & & & \tilde{\Omega}^\dagger & & & \\
     & & & & & & \ddots & & \\
     & & & & & & & \tilde{\Omega}^\dagger & \\
     & & & & & & & & \tilde{\Omega}^\dagger
    \end{array}
    \end{pmatrix}}_{\text{\LARGE$D_1$}}
    $} \\
    & \scalebox{0.53}{$
    \underbrace{\begin{pmatrix}
     I & & & & & & & & \\
     & I & & & & & & & \\
     & & I & & & & & & \\
     & & & I & & & & & \\
     & & & & I & & & & \\
     & & & & & \tilde{\Omega}^2 & & & \\
     & & & & & & \ddots & & \\
     & & & & & & & I & \\
     & & & & & & & & \tilde{\Omega}^2
    \end{pmatrix}}_{\text{\LARGE$A_2$}}
    \underbrace{\begin{pmatrix}
     I & & & & & & & & \\
     & I & & & & & & & \\
     & & I & & & & & & \\
     & & & I & & & & & \\
     & & & & \tilde{\Omega}^\dagger & & & & \\
     & & & & & \tilde{\Omega}^\dagger & & & \\
     & & & & & & \ddots & & \\
     & & & & & & & \tilde{\Omega}^\dagger & \\
     & & & & & & & & \tilde{\Omega}^\dagger
    \end{pmatrix}}_{\text{\LARGE$D_2$}}
    \hdots
    \underbrace{\begin{pmatrix}
     I & & & & & & & & \\
     & I & & & & & & & \\
     & & I & & & & & & \\
     & & & I & & & & & \\
     & & & & I & & & & \\
     & & & & & I & & & \\
     & & & & & & \ddots & & \\
     & & & & & & & I & \\
     & & & & & & & & \tilde{\Omega}^2
    \end{pmatrix}}_{\text{\LARGE$A_{k/2-1}$}}
    \underbrace{\begin{pmatrix}
     I & & & & & & & & \\
     & I & & & & & & & \\
     & & I & & & & & & \\
     & & & I & & & & & \\
     & & & & I & & & & \\
     & & & & & I & & & \\
     & & & & & & \ddots & & \\
     & & & & & & & \tilde{\Omega}^\dagger & \\
     & & & & & & & & \tilde{\Omega}^\dagger
    \end{pmatrix}}_{\text{\LARGE$D_{k/2-1}$}}
    $}.
    \label{eq:35}
\end{aligned}    
\end{equation}

This decomposition already yields substantial implementation advantages. The matrix $D_0$, containing $\tilde{\Omega}^\dagger$ on all diagonal elements, applies a phase that is uniform across all coin basis states. As a result, it is independent of the coin register, requiring no control gates, only single-qubit phase rotations on the position register. 

The matrix $A_0$, which alternates between $I$ and $\tilde{\Omega}^2$, depends solely on the state of the least significant coin qubit, $\ket{q_0^c}$. Since $\tilde{\Omega}^2$ is applied only when $\ket{q_0^c}=\ket{1}$, this operator can be implemented using single-controlled rotation gates ($C^1R_\ell)$, reducing the control overhead compared to the original $\alpha$-degree controlled gates.

\subsubsection*{Third-order decomposition and final form}
The third-order decomposition is applicable when the degree of the graph satisfies $k>4$, only on matrices $A_j$ and $D_j$ with indices $1\leq j < k/4$. At this stage, we exploit the repetitive structure of the matrices to extract common low-complexity factors, $A_0$ and $D_0$. We factor out $A_0$ and $D_0$ from the $A_j$ and $D_j$ terms, defining the residual matrices $\mathbf{A}_j$ and $\mathbf{D}_j$ as follows:
\begin{equation}
\begin{aligned}
    A_{j} &= A_0 \cdot \mathbf{A}_j, \quad \text{where} \quad \mathbf{A}_j=
    \operatorname{diag}\left(
        \underbrace{(I, \tilde{\Omega}^{\dagger2}), \dots, (I, \tilde{\Omega}^{\dagger2})}_{j}, \underbrace{I,\dots, I}_{k-2j}\right) \\[6pt]
    D_{j} &= D_0 \cdot \mathbf{D}_j, \quad \text{where} \quad \mathbf{D}_j=\operatorname{diag}\left(
    \underbrace{(\tilde{\Omega}, \tilde{\Omega}), \dots, (\tilde{\Omega}, \tilde{\Omega})}_{j}, \underbrace{I,\dots, I}_{k-2j}\right).
\end{aligned}
\label{eq:36}
\end{equation}

By grouping the common factors, we arrive at the final, highly-simplified product form of the phase shift operator:
\begin{equation}
    \Sigma = \left(A_0D_0\right)^{k/4}
    \prod_{j=1}^{k/4-1} \left(\mathbf{A}_j\mathbf{D}_j\right)\prod_{j=k/4}^{k/2-1}
    \left(A_jD_j\right).
    \label{eq:37}
\end{equation}

In explicit block matrix form, the decomposed $\Sigma$ is represented as:
\begin{equation}
\begin{aligned}
     \Sigma = & \scalebox{0.52}{${
     \underbrace{\begin{pmatrix}
     I & & & & & & & & \\
     & \tilde{\Omega}^2 & & & & & & & \\
     & & I & & & & & & \\
     & & & \tilde{\Omega}^2 & & & & & \\
     & & & & I & & & & \\
     & & & & & \tilde{\Omega}^2 & & & \\
     & & & & & & \ddots & & \\
     & & & & & & & I & \\
     & & & & & & & & \tilde{\Omega}^2
     \end{pmatrix}}_{\text{\LARGE$A_0$}}}^{\text{\large$k/4$}}{
    \underbrace{\begin{pmatrix}
     \tilde{\Omega}^\dagger & & & & & & & & \\
     & \tilde{\Omega}^\dagger & & & & & & & \\
     & & \tilde{\Omega}^\dagger & & & & & & \\
     & & & \tilde{\Omega}^\dagger & & & & & \\
     & & & & \tilde{\Omega}^\dagger & & & & \\
     & & & & & \tilde{\Omega}^\dagger & & & \\
     & & & & & & \ddots & & \\
     & & & & & & & \tilde{\Omega}^\dagger & \\
     & & & & & & & & \tilde{\Omega}^\dagger
    \end{pmatrix}}_{\text{\LARGE$D_0$}}}^{\text{\large$k/4$}}
    \underbrace{\begin{pmatrix}
     I & & & & & & & & \\
     & \tilde{\Omega}^{\dagger2} & & & & & & & \\
     & & I & & & & & & \\
     & & & I & & & & & \\
     & & & & I & & & & \\
     & & & & & I & & & \\
     & & & & & & \ddots & & \\
     & & & & & & & I & \\
     & & & & & & & & I
    \end{pmatrix}}_{\text{\LARGE$\mathbf{A}_1$}}
    \underbrace{\begin{pmatrix}
     \tilde{\Omega} & & & & & & & & \\
     & \tilde{\Omega} & & & & & & & \\
     & & I & & & & & & \\
     & & & I & & & & & \\
     & & & & I & & & & \\
     & & & & & I & & & \\
     & & & & & & \ddots & & \\
     & & & & & & & I & \\
     & & & & & & & & I
    \end{pmatrix}}_{\text{\LARGE$\mathbf{D}_1$}}
    $} \\
    & \scalebox{0.52}{$
    \underbrace{\begin{pmatrix}
     I & & & & & & & & \\
     & \tilde{\Omega}^{\dagger2} & & & & & & & \\
     & & I & & & & & & \\
     & & & \tilde{\Omega}^{\dagger2} & & & & & \\
     & & & & I & & & & \\
     & & & & & I & & & \\
     & & & & & & \ddots & & \\
     & & & & & & & I & \\
     & & & & & & & & I
    \end{pmatrix}}_{\text{\LARGE$\mathbf{A}_2$}}
    \underbrace{\begin{pmatrix}
     \tilde{\Omega} & & & & & & & & \\
     & \tilde{\Omega} & & & & & & & \\
     & & \tilde{\Omega} & & & & & & \\
     & & & \tilde{\Omega} & & & & & \\
     & & & & I & & & & \\
     & & & & & I & & & \\
     & & & & & & \ddots & & \\
     & & & & & & & I & \\
     & & & & & & & & I
    \end{pmatrix}}_{\text{\LARGE$\mathbf{D}_2$}}
    \hdots
    \underbrace{\begin{pmatrix}
     I & & & & & & & & \\
     & I & & & & & & & \\
     & & I & & & & & & \\
     & & & I & & & & & \\
     & & & & I & & & & \\
     & & & & & I & & & \\
     & & & & & & \ddots & & \\
     & & & & & & & I & \\
     & & & & & & & & \tilde{\Omega}^2
    \end{pmatrix}}_{\text{\LARGE$A_{k/2-1}$}}
    \underbrace{\begin{pmatrix}
     I & & & & & & & & \\
     & I & & & & & & & \\
     & & I & & & & & & \\
     & & & I & & & & & \\
     & & & & I & & & & \\
     & & & & & I & & & \\
     & & & & & & \ddots & & \\
     & & & & & & & \tilde{\Omega}^\dagger & \\
     & & & & & & & & \tilde{\Omega}^\dagger
    \end{pmatrix}}_{\text{\LARGE$D_{k/2-1}$}}
    $}
    \label{eq:38}
\end{aligned}    
\end{equation}

\subsection{Inverse-closed $\mathcal{S}$ with an involution}
When the generating set $\mathcal{S}$ contains a single involution, the decomposition procedure remains fundamentally similar to the non-involution case, but requires an initial adjustment to the dimensionality of the coin register. We consider a general $N$-Cayley graph with generators $\mathcal{S}=\{1, -1, 2, -2, \dots, k/2, -k/2, (k+1)/2\}$, where $\sigma_{\text{inv}}=(k+1)/2$ is the involution. 

Because the involution is its own inverse, the generating set does not include its inverse pair, yielding an odd degree $|\mathcal{S}|=k$. This poses a challenge in mapping the shift operator $S$ onto an $\alpha$-qubit coin register. To ensure compatibility with an $\alpha$-coin qubits, we apply a matrix padding by adding a diagonal identity block $I_N$ as the final block, in place of the missing inverse pair. This effectively introduces a self-loop (a step size of zero) for the last coin state, completing the dimensionality such that $k+1=2^\alpha$. 

The shift operator $S$ is diagonalized in the same manner as in the non-involution case by the SWAP-free QFT and IQFT matrices, and the modified phase shift operator $\Sigma$ becomes:
\begin{equation}
    \Sigma=\text{diag}\left(\tilde{\Omega}^\dagger, \tilde{\Omega}, \tilde{\Omega}^{\dagger2}, \tilde{\Omega}^2, \dots,\tilde{\Omega}^{\dagger (k/2)}, \tilde{\Omega}^{(k/2)}, \tilde{\Omega}^{\dagger (k+1/2)}, I\right)
\end{equation}

\subsubsection*{Hierarchical Decomposition}
The first-order decomposition on $\Sigma$ follows the same logic as the non-involution case, differing only in that the matrices $M_j$ are now defined with an expanded $(k+1)n \times (k+1)n$ dimensionality. In this step, pairs of $\tilde{\Omega}$ and $\tilde{\Omega}^\dagger$ blocks are successively factored out, resulting a product of $(k+1)/2$ $M_j$ matrices and an additional matrix $\mathcal{I}_{\text{inv}}$ that accounts for the padded identity block:
\begin{equation}
    \Sigma = \left(\prod_{j=1}^{(k-1)/2}M_j\right)\cdot \mathcal{I}_{\text{inv}}, \quad \text{where} \quad \mathcal{I}_{\text{inv}} = \left(I_N^{\oplus k} \oplus \tilde{\Omega}^{\dagger (k+1)/2} \right).
\end{equation}

The subsequent second- and third-order decompositions follow the same recursive factoring patterns for $M_j$ established in the non-involution case, while the $\mathcal{I}_{\text{inv}}$ remains intact. This results in the final product form:
\begin{equation}
    \Sigma = \left(\left(A_0D_0\right)^{(k+1)/4}
    \prod_{j=1}^{(k-3)/4} \left(\mathbf{A}_j\mathbf{D}_j\right)\prod_{j= (k+1)/4}^{(k-1)/2}(A_jD_j)\right) \cdot \mathcal{I}_{\text{inv}},
    \label{eq:40}
\end{equation}

In explicit block matrix form, the decomposed $\Sigma$ is represented as:
\begin{equation}
\begin{aligned}
     \Sigma = & \scalebox{0.48}{${
     \underbrace{\begin{pmatrix}
     I & & & & & & & & \\
     & \tilde{\Omega}^2 & & & & & & & \\
     & & I & & & & & \\
     & & & \tilde{\Omega}^2 & & & & & \\
     & & & & I & & & & \\
     & & & & & \tilde{\Omega}^2 & & & \\
     & & & & & & \ddots & & & \\
     & & & & & & & I & \\
     & & & & & & & & \tilde{\Omega}^2 \\
     \end{pmatrix}}_{\text{\LARGE$A_0$}}}^{\text{\large$(k+1)/4$}}{
    \underbrace{\begin{pmatrix}
     \tilde{\Omega}^\dagger & & & & & & & & \\
     & \tilde{\Omega}^\dagger & & & & & & & \\
     & & \tilde{\Omega}^\dagger & & & & & & \\
     & & & \tilde{\Omega}^\dagger & & & & & \\
     & & & & \tilde{\Omega}^\dagger & & & & \\
     & & & & & \tilde{\Omega}^\dagger & & & \\
     & & & & & & \ddots & & \\
     & & & & & & & \tilde{\Omega}^\dagger & \\
     & & & & & & & & \tilde{\Omega}^\dagger \\
    \end{pmatrix}}_{\text{\LARGE$D_0$}}}^{\text{\large$(k+1)/4$}}
    \underbrace{\begin{pmatrix}
     I & & & & & & & & & \\
     & \tilde{\Omega}^{\dagger2} & & & & & & & & \\
     & & I & & & & & & \\
     & & & I & & & & & \\
     & & & & I & & & & \\
     & & & & & I & & & \\
     & & & & & & \ddots & & \\
     & & & & & & & I & \\
     & & & & & & & & I \\
    \end{pmatrix}}_{\text{\LARGE$\mathbf{A}_1$}}
    \underbrace{\begin{pmatrix}
     \tilde{\Omega} & & & & & & & &\\
     & \tilde{\Omega} & & & & & & &\\
     & & I & & & & & &\\
     & & & I & & & & &\\
     & & & & I & & & &\\
     & & & & & I & & & \\
     & & & & & & \ddots & &\\
     & & & & & & & I &\\
     & & & & & & & & I\\
    \end{pmatrix}}_{\text{\LARGE$\mathbf{D}_1$}}
    $} \\
    & \scalebox{0.45}{$
    \underbrace{\begin{pmatrix}
     I & & & & & & & & \\
     & \tilde{\Omega}^{\dagger2} & & & & & & &\\
     & & I & & & & & &\\
     & & & \tilde{\Omega}^{\dagger2} & & & & &\\
     & & & & I & & & &\\
     & & & & & I & & &\\
     & & & & & & \ddots & &\\
     & & & & & & & I &\\
     & & & & & & & & I\\
    \end{pmatrix}}_{\text{\LARGE$\mathbf{A}_2$}}
    \underbrace{\begin{pmatrix}
     \tilde{\Omega} & & & & & & & & \\
     & \tilde{\Omega} & & & & & & & \\
     & & \tilde{\Omega} & & & & & & \\
     & & & \tilde{\Omega} & & & & & \\
     & & & & I & & & & \\
     & & & & & I & & & \\
     & & & & & & \ddots & & \\
     & & & & & & & I & \\
     & & & & & & & & I\\
    \end{pmatrix}}_{\text{\LARGE$\mathbf{D}_2$}}
    \hdots
    \underbrace{\begin{pmatrix}
     I & & & & & & & &\\
     & I & & & & & & &\\
     & & I & & & & & &\\
     & & & I & & & & &\\
     & & & & I & & & &\\
     & & & & & I & & &\\
     & & & & & & \ddots & &\\
     & & & & & & & I &\\
     & & & & & & & & \tilde{\Omega}^2\\
    \end{pmatrix}}_{\text{\LARGE$A_{(k-1)/2}$}}
    \underbrace{\begin{pmatrix}
     I & & & & & & & &\\
     & I & & & & & & &\\
     & & I & & & & & &\\
     & & & I & & & & &\\
     & & & & I & & & &\\
     & & & & & I & & &\\
     & & & & & & \ddots & &\\
     & & & & & & & \tilde{\Omega}^\dagger &\\
     & & & & & & & & \tilde{\Omega}^\dagger\\
    \end{pmatrix}}_{\text{\LARGE$D_{(k-1)/2}$}}
    \underbrace{\begin{pmatrix}
     I & & & & & & & & \\
     & I & & & & & & & \\
     & & I & & & & & & \\
     & & & I & & & & & \\
     & & & & I & & & & \\
     & & & & & I & & & \\
     & & & & & & \ddots & &\\
     & & & & & & & I & \\
     & & & & & & & & \tilde{\Omega}^{\dagger (k+1)/2}\\
    \end{pmatrix}}_{\text{\LARGE $\mathcal{I}_{\text{inv}}$}}
    $}
    \label{eq:42}
\end{aligned}    
\end{equation}

\subsection{Non-inverse-closed $\mathcal{S}$}
When the generating set is not closed under inverses, $\mathcal{S}$ contains no inverse elements. Consequently, the resulting $N$-Cayley graph $\Gamma\left(\mathbb{Z}_N, \{1, 2, \dots, k\} \right)$ is directed, and the shift operator $S$ contains strictly of powers of the increment operator $P_0$, lacking the alternating $P_0$ and $P_0^\mathsf{T}$ structure found in undirected graphs. Following the same diagonalization procedure using the SWAP-free QFT matrices, the phase shift operator $\Sigma$ takes the form:
\begin{equation}
    \Sigma = \begin{pmatrix}
        \tilde{\Omega}^\dagger&&&&&&&\\
        &\tilde{\Omega}^{\dagger2}&&&&&&\\
        &&\tilde{\Omega}^{\dagger3}&&&&&\\
        &&&\tilde{\Omega}^{\dagger4}&&&&\\
        &&&&\tilde{\Omega}^{\dagger5}&&&\\
        &&&&&\tilde{\Omega}^{\dagger6}&&\\
        &&&&&&\ddots&\\
        &&&&&&&\tilde{\Omega}^{\dagger k}\\
    \end{pmatrix}
\end{equation}

Unlike the inverse-closed case, the decomposition of $\Sigma$ is considerably simplified. Since every diagonal block is a power of the same base operator $\tilde{\Omega}^\dagger$, we factor it out sequentially using a method analogous to our first-order decomposition:
\begin{equation}
    \Sigma = \prod_{j=0}^{k-1}D_j \quad \text{where} \quad D_j = \text{diag}\left(\underbrace{I, \dots, I}_{j}, \underbrace{\tilde{\Omega}^\dagger, \dots \tilde{\Omega}^\dagger}_{k-j}\right). 
    \label{eq:D_j_non-inverse-closed}
\end{equation}
Explicitly, this product takes the form:
\begin{equation}
    \begin{aligned}
    \Sigma = \scalebox{0.53}{$\underbrace{\begin{pmatrix}
     \tilde{\Omega}^\dagger & & & & & & & & \\
     & \tilde{\Omega}^\dagger & & & & & & & \\
     & & \tilde{\Omega}^\dagger & & & & & & \\
     & & & \tilde{\Omega}^\dagger & & & & & \\
     & & & & \tilde{\Omega}^\dagger & & & & \\
     & & & & & \tilde{\Omega}^\dagger & & & \\
     & & & & & & \ddots & & \\
     & & & & & & & \tilde{\Omega}^\dagger & \\
     & & & & & & & & \tilde{\Omega}^\dagger
     \end{pmatrix}}_{\text{\LARGE$D_0$}}\underbrace{\begin{pmatrix}
     I & & & & & & & & \\
     & \tilde{\Omega}^\dagger & & & & & & & \\
     & & \tilde{\Omega}^\dagger & & & & & & \\
     & & & \tilde{\Omega}^\dagger & & & & & \\
     & & & & \tilde{\Omega}^\dagger & & & & \\
     & & & & & \tilde{\Omega}^\dagger & & & \\
     & & & & & & \ddots & & \\
     & & & & & & & \tilde{\Omega}^\dagger & \\
     & & & & & & & & \tilde{\Omega}^\dagger
     \end{pmatrix}}_{\text{\LARGE$D_1$}}\hdots\underbrace{\begin{pmatrix}
     I & & & & & & & & \\
     & I & & & & & & & \\
     & & I & & & & & & \\
     & & & I & & & & & \\
     & & & & I & & & & \\
     & & & & & I & & & \\
     & & & & & & \ddots & & \\
     & & & & & & & \tilde{\Omega}^\dagger & \\
     & & & & & & & & \tilde{\Omega}^\dagger
     \end{pmatrix}}_{\text{\LARGE$D_{k-2}$}}\underbrace{\begin{pmatrix}
     I & & & & & & & & \\
     & I & & & & & & & \\
     & & I & & & & & & \\
     & & & I & & & & & \\
     & & & & I & & & & \\
     & & & & & I & & & \\
     & & & & & & \ddots & & \\
     & & & & & & & I & \\
     & & & & & & & & \tilde{\Omega}^\dagger
    \end{pmatrix}}_{\text{\LARGE$D_{k-1}$}}$}.
    \end{aligned}
    \label{eq:43}
\end{equation}

Note that the index $j$ for $D_j$ is defined differently here (Eq.~\ref{eq:D_j_non-inverse-closed}) than in the inverse-closed case (Eq.~\ref{eq:36}). Notably, because $D_j$ consists only of uniform $\tilde{\Omega}^\dagger$, the second-order decomposition required in the inverse-closed case to handle ($\tilde{\Omega}^\dagger,\tilde{\Omega}$) blocks is entirely bypassed.

For indices $1 \leq j < k/2$ ($k$ must be greater than 2), the matrices $D_j$ admit a further decomposition analogous to the third-order decomposition in the inverse-closed case, factoring out the global $D_0$ term to define residual matrices $\mathbf{D}_j$:
\begin{equation}
    D_{1\leq j< k/4} = D_0 \cdot \mathbf{D}_j \quad \text{where} \quad \mathbf{D}_j=\operatorname{diag}\left(
        \underbrace{\tilde{\Omega}, \dots, \tilde{\Omega}}_{j}, \underbrace{I,\dots, I}_{k-j}\right).
\end{equation}

The fully decomposed phase shift operator $\Sigma$ is expressed as:
\begin{equation}
    \Sigma = D_0^{ k/2} \prod_{j=1}^{k/2-1} \mathbf{D}_j \prod_{j=k/2}^{k-1} D_j.
\end{equation}
The explicit block matrix expansion is given by:
\begin{equation}
    \begin{aligned}
    \Sigma = \scalebox{0.53}{${\underbrace{\begin{pmatrix}
     \tilde{\Omega}^\dagger & & & & & & & & \\
     & \tilde{\Omega}^\dagger & & & & & & & \\
     & & \tilde{\Omega}^\dagger & & & & & & \\
     & & & \tilde{\Omega}^\dagger & & & & & \\
     & & & & \tilde{\Omega}^\dagger & & & & \\
     & & & & & \tilde{\Omega}^\dagger & & & \\
     & & & & & & \ddots & & \\
     & & & & & & & \tilde{\Omega}^\dagger & \\
     & & & & & & & & \tilde{\Omega}^\dagger
     \end{pmatrix}}_{\text{\LARGE$D_0$}}}^{\text{\large$ k/2$}}\underbrace{\begin{pmatrix}
     \tilde{\Omega} & & & & & & & & \\
     & I & & & & & & & \\
     & & I & & & & & & \\
     & & & I & & & & & \\
     & & & & I & & & & \\
     & & & & & I & & & \\
     & & & & & & \ddots & & \\
     & & & & & & & I & \\
     & & & & & & & & I \end{pmatrix}}_{\text{\LARGE$\mathbf{D}_1$}}\hdots\underbrace{\begin{pmatrix}
     I & & & & & & & & \\
     & I & & & & & & & \\
     & & I & & & & & & \\
     & & & I & & & & & \\
     & & & & I & & & & \\
     & & & & & I & & & \\
     & & & & & & \ddots & & \\
     & & & & & & & \Omega^\dagger & \\
     & & & & & & & & \Omega^\dagger
     \end{pmatrix}}_{\text{\LARGE$D_{k-2}$}}\underbrace{\begin{pmatrix}
     I & & & & & & & & \\
     & I & & & & & & & \\
     & & I & & & & & & \\
     & & & I & & & & & \\
     & & & & I & & & & \\
     & & & & & I & & & \\
     & & & & & & \ddots & & \\
     & & & & & & & I & \\
     & & & & & & & & \Omega^\dagger
    \end{pmatrix}}_{\text{\LARGE$D_{k-1}$}}$}.
    \end{aligned}
    \label{eq:46}
\end{equation}

The implementation of the matrices $D_0$, $D_j$, and $\mathbf{D}_j$ is equivalent to that of the inverse-closed case. A comprehensive analysis of the resulting gate complexity is provided in the following section.

\subsection{Final gate complexity analysis}
\label{sec:gate_complexity}
The systematic decomposition of the diagonal phase shift operator $\Sigma$ into a product of simpler factors significantly enhances the efficiency of the quantum circuit. by leveraging block-wise factorization, we replace high-degree multi-qubit controlled gates with a streamlined set of lower-degree operations. In this section, we quantify the resources required for each factor ($A_j, D_j, \mathbf{A}_j$, $\mathbf{D}_j$ and $\mathcal{I}_{\text{inv}}$), demonstrating the reduction in control logic overhead. 

\subsubsection*{Primary factors: $A_0$ and $D_0$}  
\paragraph{The $A_0$ operator} As established in Eq.~\ref{eq:38}, $A_0$ is raised to $k/4$ (non-involution), or $(k+1)/4$ (involution). It implements the diagonal phase block $\tilde{\Omega}^{2^{\alpha-1}}$ conditionally, based only on the state of the least significant coin qubit, $\ket{q_0^c}=\ket{1}$. Extending the construction in Eq.~\ref{eq:24}, rotation gates required to implement any diagonal phase operator raised to a power of $2^p$ can be generalized as:
\begin{equation}
    \tilde{\Omega}^{2^p} = \bigotimes_{\ell=n-1}^{0}R_{\ell+1}^{2^p} = \bigotimes_{\ell=n-1}^0 R_{\ell+1-p} = R_{n-p} \otimes R_{n-1-p} \otimes \cdots \otimes I \otimes \cdots \otimes R_{1-p}.
\end{equation}
Because this operation is controlled by only one coin qubit, it requires $n-1$ single-controlled gates, $C^{(1)}R_{\ell+2-\alpha}$ to implement. 

\paragraph{The $D_0$ operator} Similarly, $D_0$ (raised to the same powers) is  composed of uniform $\tilde{\Omega}$ blocks across all coin states. Since its implementation is entirely coin-independent, it only requires no control logic. It is implemented using $n-1$ single-qubit rotation gates, $R^\dagger_{\ell+3-\alpha}$ on the position register.

\subsubsection*{Specialized Blocks and Residual Matrices}
\paragraph{Involution block matrix $\mathcal{I}_{\text{inv}}$} In cases involving an involution, the matrix $\mathcal{I}_{\text{inv}}$ can be fused with $A_{(k-1)/2}$ from Eq.~\ref{eq:42} to achieve the form $\left(I^{\oplus k} \oplus \tilde{\Omega}^{\dagger (k-3)/2} \right)$ since they both are conditioned on the same coin state. This combined operator depends exclusively on the final coin state $\ket{k-1}$, necessitating $n$ multi-controlled gates of degree $\alpha$, $C^{(\alpha)}R_{\ell+1}$. 

\paragraph{Residual matrices ($A_j, D_j, \mathbf{A}_j, \mathbf{D}_j$)} The complexity of the remaining factors depends on the number of active operations (non-identity blocks, such as $\tilde{\Omega}$, $\tilde{\Omega}^\dagger$, or their powers) that are conditioned on the coin states, denoted by $m$. The degree of the required controlled gates is determined by the binary representation of $m$ through two types of binary decomposition. 
\begin{itemize}
    \item \textbf{Additive Decomposition:} A standard binary decomposition $m = \sum_{e\in \mathbb{N}} 2^e$ dictates the control logic. Each term $2^e$ requires ($\alpha-e$)-degree controlled gates ($C^{(\alpha-e)}R_\lambda$). For example, for $m=3$ (as seen in $A_3$ or directed $D_3$ case), the additive form $3=2^1+2^0$ requires $C^{(\alpha-1)}R_\lambda$ and $C^{(\alpha)}R_\lambda$ gates.

    \item \textbf{Subtractive Decomposition:} To further optimize the control operation, we employ subtractive binary decomposition, representing $m$ as a combination of sum and differences powers: $m=\sum_{e\in \mathbb{N}} 2^e-\sum_{e\in \mathbb{N}} 2^e$, where $e \leq \alpha$. Under this approach, the configuration $m=3$ can be optimized as $3=2^2-2^0$, which allows us to substitute higher-degree controlled gates with lower-degree alternatives. This would necessitate $C^{(\alpha-2)}R_\lambda$ and $C^{(\alpha)}R_\lambda^\dagger$ gates.
\end{itemize}
The additive decomposition only provides a naive upper bound on the control logic requirements. To further optimize the circuit, the key is to leverage the fact that $m$ can also be represented as differences of powers through subtractive decomposition. This distinction is crucial because it enables the replacement of higher-degree controlled gates with lower-degree controlled gates, while maintaining scaling complexity. This strategy dictates the control logic: we first apply the phase rotation to a broader set of coin states using lower-degree controlled gates (e.g., $C^{(\alpha-2)}R_\lambda$), followed by the inverse operation to "undo" the phase on specific coin states using higher-degree controlled gates (e.g., $C^{(\alpha)}R_\lambda$). By shifting the operations to lower-degree controls, we reduce the average control overhead across the circuit.

\subsection{Example: DTQW on an 8-Cayley Graph with an Involution}
To illustrate the implementation procedure, we consider a DTQW on the group $\mathbb{Z}_8$ with the generating set $\mathcal{S}= \{1,-1,2,-2,3,-3,4\}$. 
This yields a Cayley graph $\Gamma\left(\mathbb{Z}_8, \mathcal{S}\right)$ of degree $k=7$ containing an involution, $4 \equiv 4 \pmod 8$. In this section, we provide the explicit decomposition of the shift operator, the corresponding circuit diagram, and an analysis on gate complexities required for its implementation.

The $k=7$ generators, each associated with a specific coin state, are encoded into $\alpha=3$ coin qubits. We employ matrix padding to leave the last basis state $\ket{7}$ as an identity (representing a self-loop), while the  $N=8$ position states are encoded into $n=3$ position qubits. 

The shift operator $S$ is defined a $64 \times 64$ block diagonal matrix:
\begin{equation}
    S = \begin{pmatrix}
        P_0 & & & & & & & \\
        & P_0^{\mathsf{T}} & & & & & & \\
        & & P_0^{2} & & & & & \\
        & & & P_0^{\mathsf{T}2} & & & & \\
        & & & & P_0^{3} & & & \\
        & & & & & P_0^{\mathsf{T}3} & & \\
        & & & & & & P_0^{4} & \\
        & & & & & & & I \\
    \end{pmatrix},
\end{equation}
where each ID block is of size $8 \times 8$. We diagonalize $S$ using the SWAP-free QFT ($I_8\otimes \tilde{\mathcal{F}}$) and its inverse ($I_8\otimes \tilde{\mathcal{F}}^\dagger$), yielding the phase shift operator:
\begin{equation}
    \Sigma = \begin{pmatrix}
        \tilde{\Omega}^{\dagger} & & & & & & & \\
        & \tilde{\Omega} & & & & & & \\
        & & \tilde{\Omega}^{\dagger2} & & & & & \\
        & & & \tilde{\Omega}^2 & & & & \\
        & & & & \tilde{\Omega}^{\dagger3} & & & \\
        & & & & & \tilde{\Omega}^3 & & \\
        & & & & & & \tilde{\Omega}^{\dagger4} & \\
        & & & & & & & I \\
        \end{pmatrix}.
\end{equation}
The block $\tilde{\Omega}^{\dagger4}$ corresponds to the involution and is therefore not paired with an inverse block. We follow the decomposition method established in the previous section for the involution case. 

\subsubsection{Decomposition Stages}
\paragraph{First-order}
We first isolate the involution block $\mathcal{I}_{\text{inv}}$ and factor the remaining paired blocks into four $M_j$ matrices. This yields:
\begin{equation}
\begin{gathered}
    \Sigma= M_0 \cdot M_1\cdot M_2 \cdot M_3\cdot\mathcal{I}_{\text{inv}} \\[6pt]
    = \scalebox{0.58}{$\begin{pmatrix}
        \begin{array}{@{\hskip 2.8pt}c@{\hskip 2.8pt} @{\hskip 2.8pt}c@{\hskip 2.8pt} @{\hskip 2.8pt}c@{\hskip 2.8pt} @{\hskip 2.8pt}c@{\hskip 2.8pt} @{\hskip 2.8pt}c@{\hskip 2.8pt} @{\hskip 2.8pt}c@{\hskip 2.8pt} @{\hskip 2.8pt}c@{\hskip 2.8pt} @{\hskip 2.8pt}c@{\hskip 2.8pt}}
             \tilde{\Omega}^{\dagger} & & & & & & & \\
        & \tilde{\Omega} & & & & & & \\
        & & \tilde{\Omega}^{\dagger} & & & & & \\
        & & & \tilde{\Omega} & & & & \\
        & & & & \tilde{\Omega}^{\dagger} & & & \\
        & & & & & \tilde{\Omega} & & \\
        & & & & & & \tilde{\Omega}^{\dagger} & \\
        & & & & & & & \tilde{\Omega}
        \end{array}
        \end{pmatrix}\begin{pmatrix}
        \begin{array}{@{\hskip 3.3pt}c@{\hskip 3.3pt} @{\hskip 3.3pt}c@{\hskip 3.3pt} @{\hskip 3.3pt}c@{\hskip 3.3pt} @{\hskip 3.3pt}c@{\hskip 3.3pt} @{\hskip 3.3pt}c@{\hskip 3.3pt} @{\hskip 3.3pt}c@{\hskip 3.3pt} @{\hskip 3.3pt}c@{\hskip 3.3pt} @{\hskip 3.3pt}c@{\hskip 3.3pt}}
        I & & & & & & & \\
        & I & & & & & & \\
        & & \tilde{\Omega}^{\dagger} & & & & & \\
        & & & \tilde{\Omega} & & & & \\
        & & & & \tilde{\Omega}^{\dagger} & & & \\
        & & & & & \tilde{\Omega} & & \\
        & & & & & & \tilde{\Omega}^{\dagger} & \\
        & & & & & & & \tilde{\Omega}
        \end{array}
        \end{pmatrix}\begin{pmatrix}
        \begin{array}{@{\hskip 3.5pt}c@{\hskip 3.5pt} @{\hskip 3.5pt}c@{\hskip 3.5pt} @{\hskip 3.5pt}c@{\hskip 3.5pt} @{\hskip 3.5pt}c@{\hskip 3.5pt} @{\hskip 3.5pt}c@{\hskip 3.5pt} @{\hskip 3.5pt}c@{\hskip 3.5pt} @{\hskip 3.5pt}c@{\hskip 3.5pt} @{\hskip 3.5pt}c@{\hskip 3.5pt}}
             I & & & & & & & \\
        & I & & & & & & \\
        & & I & & & & & \\
        & & & I & & & & \\
        & & & & \tilde{\Omega}^{\dagger} & & & \\
        & & & & & \tilde{\Omega} & & \\
        & & & & & & \tilde{\Omega}^{\dagger} & \\
        & & & & & & & \tilde{\Omega}
        \end{array}
        \end{pmatrix}\begin{pmatrix}
        \begin{array}{@{\hskip 3.6pt}c@{\hskip 3.6pt} @{\hskip 3.6pt}c@{\hskip 3.6pt} @{\hskip 3.6pt}c@{\hskip 3.6pt} @{\hskip 3.6pt}c@{\hskip 3.6pt} @{\hskip 3.6pt}c@{\hskip 3.6pt} @{\hskip 3.6pt}c@{\hskip 3.6pt} @{\hskip 3.6pt}c@{\hskip 3.6pt} @{\hskip 3.6pt}c@{\hskip 3.6pt}}
             I & & & & & & & \\
        & I & & & & & & \\
        & & I & & & & & \\
        & & & I & & & & \\
        & & & & I & & & \\
        & & & & & I & & \\
        & & & & & & \tilde{\Omega}^{\dagger} & \\
        & & & & & & & \tilde{\Omega}
        \end{array}
        \end{pmatrix}\begin{pmatrix}
        \begin{array}{@{\hskip 3.8pt}c@{\hskip 3.8pt} @{\hskip 3.8pt}c@{\hskip 3.8pt} @{\hskip 3.8pt}c@{\hskip 3.8pt} @{\hskip 3.8pt}c@{\hskip 3.8pt} @{\hskip 3.8pt}c@{\hskip 3.8pt} @{\hskip 3.8pt}c@{\hskip 3.8pt} @{\hskip 3.8pt}c@{\hskip 3.8pt} @{\hskip 3.8pt}c@{\hskip 3.8pt}}
             I & & & & & & & \\
        & I & & & & & & \\
        & & I & & & & & \\
        & & & I & & & & \\
        & & & & I & & & \\
        & & & & & I & & \\
        & & & & & & I & \\
        & & & & & & & \tilde{\Omega}^{\dagger4}
        \end{array}
        \end{pmatrix}$}.
\end{gathered}
\end{equation}

\paragraph{Second-order}
Each $M_j$ is further decomposed into $A_j$ and $D_j$ pairs. This step separates the alternating phase directions while leaving the involution block $\mathcal{I}_{\text{inv}}$ intact. The operator $\Sigma$ is then expressed as:
\begin{equation}
\begin{gathered}
    \Sigma = A_0 D_0\cdot A_1D_1\cdot A_2D_2 \cdot A_3D_3\cdot\mathcal{I}_{\text{inv}} \\[6pt]
    = \scalebox{0.69}{$
    \begin{pmatrix}
    \begin{array}{@{\hskip 3.2pt}c@{\hskip 3.2pt} @{\hskip 3.2pt}c@{\hskip 3.2pt} @{\hskip 3.2pt}c@{\hskip 3.2pt} @{\hskip 3.2pt}c@{\hskip 3.2pt} @{\hskip 3.2pt}c@{\hskip 3.2pt} @{\hskip 3.2pt}c@{\hskip 3.2pt} @{\hskip 3.2pt}c@{\hskip 3.2pt} @{\hskip 3.2pt}c@{\hskip 3.2pt}}
         I &&&&&&& \\
        &\tilde{\Omega}^2 &&&&&& \\
        &&I&&&&& \\
        &&&\tilde{\Omega}^2&&&& \\
        &&&&I&&& \\
        &&&&&\tilde{\Omega}^2&& \\
        &&&&&&I& \\
        &&&&&&&\tilde{\Omega}^2 \\
    \end{array}
    \end{pmatrix}\begin{pmatrix}
    \begin{array}{@{\hskip 2.6pt}c@{\hskip 2.6pt} @{\hskip 2.6pt}c@{\hskip 2.6pt} @{\hskip 2.6pt}c@{\hskip 2.6pt} @{\hskip 2.6pt}c@{\hskip 2.6pt} @{\hskip 2.6pt}c@{\hskip 2.6pt} @{\hskip 2.6pt}c@{\hskip 2.6pt} @{\hskip 2.6pt}c@{\hskip 2.6pt} @{\hskip 2.6pt}c@{\hskip 2.6pt}}
        \tilde{\Omega}^\dagger&&&&&&& \\
        &\tilde{\Omega}^\dagger &&&&&& \\
        &&\tilde{\Omega}^\dagger&&&&& \\
        &&&\tilde{\Omega}^\dagger&&&& \\
        &&&&\tilde{\Omega}^\dagger&&& \\
        &&&&&\tilde{\Omega}^\dagger&& \\
        &&&&&&\tilde{\Omega}^\dagger& \\
        &&&&&&&\tilde{\Omega}^\dagger \\
    \end{array}
    \end{pmatrix}\begin{pmatrix}
    \begin{array}{@{\hskip 3.2pt}c@{\hskip 3.2pt} @{\hskip 3.2pt}c@{\hskip 3.2pt} @{\hskip 3.2pt}c@{\hskip 3.2pt} @{\hskip 3.2pt}c@{\hskip 3.2pt} @{\hskip 3.2pt}c@{\hskip 3.2pt} @{\hskip 3.2pt}c@{\hskip 3.2pt} @{\hskip 3.2pt}c@{\hskip 3.2pt} @{\hskip 3.2pt}c@{\hskip 3.2pt}}
        I &&&&&&& \\
        &I &&&&&& \\
        &&I&&&&& \\
        &&&\tilde{\Omega}^2&&&& \\
        &&&&I&&& \\
        &&&&&\tilde{\Omega}^2&& \\
        &&&&&&I& \\
        &&&&&&&\tilde{\Omega}^2 \\
    \end{array}
    \end{pmatrix}\begin{pmatrix}
    \begin{array}{@{\hskip 2.6pt}c@{\hskip 2.6pt} @{\hskip 2.6pt}c@{\hskip 2.6pt} @{\hskip 2.6pt}c@{\hskip 2.6pt} @{\hskip 2.6pt}c@{\hskip 2.6pt} @{\hskip 2.6pt}c@{\hskip 2.6pt} @{\hskip 2.6pt}c@{\hskip 2.6pt} @{\hskip 2.6pt}c@{\hskip 2.6pt} @{\hskip 2.6pt}c@{\hskip 2.6pt}}
        I&&&&&&& \\
        &I&&&&&& \\
        &&\tilde{\Omega}^\dagger&&&&& \\
        &&&\tilde{\Omega}^\dagger&&&& \\
        &&&&\tilde{\Omega}^\dagger&&& \\
        &&&&&\tilde{\Omega}^\dagger&& \\
        &&&&&&\tilde{\Omega}^\dagger& \\
        &&&&&&&\tilde{\Omega}^\dagger \\
    \end{array}
    \end{pmatrix}$} \\
    \scalebox{0.595}{$
    \begin{pmatrix}
    \begin{array}{@{\hskip 3.4pt}c@{\hskip 3.4pt} @{\hskip 3.4pt}c@{\hskip 3.4pt} @{\hskip 3.4pt}c@{\hskip 3.4pt} @{\hskip 3.4pt}c@{\hskip 3.4pt} @{\hskip 3.4pt}c@{\hskip 3.4pt} @{\hskip 3.4pt}c@{\hskip 3.4pt} @{\hskip 3.4pt}c@{\hskip 3.4pt} @{\hskip 3.4pt}c@{\hskip 3.4pt}}
         I &&&&&&& \\
        &I &&&&&& \\
        &&I&&&&& \\
        &&&I&&&& \\
        &&&&I&&& \\
        &&&&&\tilde{\Omega}^2&& \\
        &&&&&&I& \\
        &&&&&&&\tilde{\Omega}^2 \\
    \end{array}
    \end{pmatrix}\begin{pmatrix}
    \begin{array}{@{\hskip 3.6pt}c@{\hskip 3.6pt} @{\hskip 3.6pt}c@{\hskip 3.6pt} @{\hskip 3.6pt}c@{\hskip 3.6pt} @{\hskip 3.6pt}c@{\hskip 3.6pt} @{\hskip 3.6pt}c@{\hskip 3.6pt} @{\hskip 3.6pt}c@{\hskip 3.6pt} @{\hskip 3.6pt}c@{\hskip 3.6pt} @{\hskip 3.6pt}c@{\hskip 3.6pt}}
        I&&&&&&& \\
        &I&&&&&& \\
        &&I&&&&& \\
        &&&I&&&& \\
        &&&&\tilde{\Omega}^\dagger&&& \\
        &&&&&\tilde{\Omega}^\dagger&& \\
        &&&&&&\tilde{\Omega}^\dagger& \\
        &&&&&&&\tilde{\Omega}^\dagger \\
    \end{array}
    \end{pmatrix}\begin{pmatrix}
    \begin{array}{@{\hskip 3.8pt}c@{\hskip 3.8pt} @{\hskip 3.8pt}c@{\hskip 3.8pt} @{\hskip 3.8pt}c@{\hskip 3.8pt} @{\hskip 3.8pt}c@{\hskip 3.8pt} @{\hskip 3.8pt}c@{\hskip 3.8pt} @{\hskip 3.8pt}c@{\hskip 3.8pt} @{\hskip 3.8pt}c@{\hskip 3.8pt} @{\hskip 3.8pt}c@{\hskip 3.8pt}}
        I &&&&&&& \\
        &I &&&&&& \\
        &&I&&&&& \\
        &&&I&&&& \\
        &&&&I&&& \\
        &&&&&I&& \\
        &&&&&&I& \\
        &&&&&&&\tilde{\Omega}^2 \\
    \end{array}
    \end{pmatrix}\begin{pmatrix}
    \begin{array}{@{\hskip 3.6pt}c@{\hskip 3.6pt} @{\hskip 3.6pt}c@{\hskip 3.6pt} @{\hskip 3.6pt}c@{\hskip 3.6pt} @{\hskip 3.6pt}c@{\hskip 3.6pt} @{\hskip 3.6pt}c@{\hskip 3.6pt} @{\hskip 3.6pt}c@{\hskip 3.6pt} @{\hskip 3.6pt}c@{\hskip 3.6pt} @{\hskip 3.6pt}c@{\hskip 3.6pt}}
        I&&&&&&& \\
        &I&&&&&& \\
        &&I&&&&& \\
        &&&I&&&& \\
        &&&&I&&& \\
        &&&&&I&& \\
        &&&&&&\tilde{\Omega}^\dagger& \\
        &&&&&&&\tilde{\Omega}^\dagger \\
    \end{array}
    \end{pmatrix}\begin{pmatrix}
    \begin{array}{@{\hskip 3.8pt}c@{\hskip 3.8pt} @{\hskip 3.8pt}c@{\hskip 3.8pt} @{\hskip 3.8pt}c@{\hskip 3.8pt} @{\hskip 3.8pt}c@{\hskip 3.8pt} @{\hskip 3.8pt}c@{\hskip 3.8pt} @{\hskip 3.8pt}c@{\hskip 3.8pt} @{\hskip 3.8pt}c@{\hskip 3.8pt} @{\hskip 3.8pt}c@{\hskip 3.8pt}}
        I&&&&&&& \\
        &I&&&&&& \\
        &&I&&&&& \\
        &&&I&&&& \\
        &&&&I&&& \\
        &&&&&I&& \\
        &&&&&&I& \\
        &&&&&&&\tilde{\Omega}^{\dagger 4} \\
    \end{array}
    \end{pmatrix}$}.
\end{gathered}
\end{equation}

\paragraph{Third-order}
The final decomposition applies only to the $A_1$ and $D_1$ matrices, where $A_0$ and $D_0$ are factored out leaving the residual forms $\mathbf{A}_1$ and $\mathbf{D}_1$. This yields the final decomposed form:
\begin{equation}
    \begin{gathered}
    \Sigma = (A_0 D_0)^{2}\cdot \mathbf{A}_1 \mathbf{D}_1\cdot A_2D_2 \cdot A_3D_3\cdot\mathcal{I}_{\text{inv}} \\[6pt]
    = \scalebox{0.72}{$
    \begin{pmatrix}
    \begin{array}{@{\hskip 3.2pt}c@{\hskip 3.2pt} @{\hskip 3.2pt}c@{\hskip 3.2pt} @{\hskip 3.2pt}c@{\hskip 3.2pt} @{\hskip 3.2pt}c@{\hskip 3.2pt} @{\hskip 3.2pt}c@{\hskip 3.2pt} @{\hskip 3.2pt}c@{\hskip 3.2pt} @{\hskip 3.2pt}c@{\hskip 3.2pt} @{\hskip 3.2pt}c@{\hskip 3.2pt}}
         I &&&&&&& \\
        &\tilde{\Omega}^2 &&&&&& \\
        &&I&&&&& \\
        &&&\tilde{\Omega}^2&&&& \\
        &&&&I&&& \\
        &&&&&\tilde{\Omega}^2&& \\
        &&&&&&I& \\
        &&&&&&&\tilde{\Omega}^2 \\
    \end{array}
    \end{pmatrix}^{\text{\large 2}}\begin{pmatrix}
    \begin{array}{@{\hskip 2.6pt}c@{\hskip 2.6pt} @{\hskip 2.6pt}c@{\hskip 2.6pt} @{\hskip 2.6pt}c@{\hskip 2.6pt} @{\hskip 2.6pt}c@{\hskip 2.6pt} @{\hskip 2.6pt}c@{\hskip 2.6pt} @{\hskip 2.6pt}c@{\hskip 2.6pt} @{\hskip 2.6pt}c@{\hskip 2.6pt} @{\hskip 2.6pt}c@{\hskip 2.6pt}}
        \tilde{\Omega}^\dagger&&&&&&& \\
        &\tilde{\Omega}^\dagger &&&&&& \\
        &&\tilde{\Omega}^\dagger&&&&& \\
        &&&\tilde{\Omega}^\dagger&&&& \\
        &&&&\tilde{\Omega}^\dagger&&& \\
        &&&&&\tilde{\Omega}^\dagger&& \\
        &&&&&&\tilde{\Omega}^\dagger& \\
        &&&&&&&\tilde{\Omega}^\dagger \\
    \end{array}
    \end{pmatrix}^{\text{\large 2}}\begin{pmatrix}
    \begin{array}{@{\hskip 3.2pt}c@{\hskip 3.2pt} @{\hskip 3.2pt}c@{\hskip 3.2pt} @{\hskip 3.2pt}c@{\hskip 3.2pt} @{\hskip 3.2pt}c@{\hskip 3.2pt} @{\hskip 3.2pt}c@{\hskip 3.2pt} @{\hskip 3.2pt}c@{\hskip 3.2pt} @{\hskip 3.2pt}c@{\hskip 3.2pt} @{\hskip 3.2pt}c@{\hskip 3.2pt}}
        I &&&&&&& \\
        &\tilde{\Omega}^{\dagger2}&&&&&& \\
        &&I&&&&& \\
        &&&I&&&& \\
        &&&&I&&& \\
        &&&&&I&& \\
        &&&&&&I& \\
        &&&&&&&I \\
    \end{array}
    \end{pmatrix}\begin{pmatrix}
    \begin{array}{@{\hskip 3.6pt}c@{\hskip 3.6pt} @{\hskip 3.6pt}c@{\hskip 3.6pt} @{\hskip 3.6pt}c@{\hskip 3.6pt} @{\hskip 3.6pt}c@{\hskip 3.6pt} @{\hskip 3.6pt}c@{\hskip 3.6pt} @{\hskip 3.6pt}c@{\hskip 3.6pt} @{\hskip 3.6pt}c@{\hskip 3.6pt} @{\hskip 3.6pt}c@{\hskip 3.6pt}}
        \tilde{\Omega}&&&&&&& \\
        &\tilde{\Omega}&&&&&& \\
        &&I&&&&& \\
        &&&I&&&& \\
        &&&&I&&& \\
        &&&&&I&& \\
        &&&&&&I& \\
        &&&&&&&I \\
    \end{array}
    \end{pmatrix}$} \\
    \scalebox{0.61}{$
    \begin{pmatrix}
    \begin{array}{@{\hskip 3.4pt}c@{\hskip 3.4pt} @{\hskip 3.4pt}c@{\hskip 3.4pt} @{\hskip 3.4pt}c@{\hskip 3.4pt} @{\hskip 3.4pt}c@{\hskip 3.4pt} @{\hskip 3.4pt}c@{\hskip 3.4pt} @{\hskip 3.4pt}c@{\hskip 3.4pt} @{\hskip 3.4pt}c@{\hskip 3.4pt} @{\hskip 3.4pt}c@{\hskip 3.4pt}}
         I &&&&&&& \\
        &I &&&&&& \\
        &&I&&&&& \\
        &&&I&&&& \\
        &&&&I&&& \\
        &&&&&\tilde{\Omega}^2&& \\
        &&&&&&I& \\
        &&&&&&&\tilde{\Omega}^2 \\
    \end{array}
    \end{pmatrix}\begin{pmatrix}
    \begin{array}{@{\hskip 3.6pt}c@{\hskip 3.6pt} @{\hskip 3.6pt}c@{\hskip 3.6pt} @{\hskip 3.6pt}c@{\hskip 3.6pt} @{\hskip 3.6pt}c@{\hskip 3.6pt} @{\hskip 3.6pt}c@{\hskip 3.6pt} @{\hskip 3.6pt}c@{\hskip 3.6pt} @{\hskip 3.6pt}c@{\hskip 3.6pt} @{\hskip 3.6pt}c@{\hskip 3.6pt}}
        I&&&&&&& \\
        &I&&&&&& \\
        &&I&&&&& \\
        &&&I&&&& \\
        &&&&\tilde{\Omega}^\dagger&&& \\
        &&&&&\tilde{\Omega}^\dagger&& \\
        &&&&&&\tilde{\Omega}^\dagger& \\
        &&&&&&&\tilde{\Omega}^\dagger \\
    \end{array}
    \end{pmatrix}\begin{pmatrix}
    \begin{array}{@{\hskip 3.8pt}c@{\hskip 3.8pt} @{\hskip 3.8pt}c@{\hskip 3.8pt} @{\hskip 3.8pt}c@{\hskip 3.8pt} @{\hskip 3.8pt}c@{\hskip 3.8pt} @{\hskip 3.8pt}c@{\hskip 3.8pt} @{\hskip 3.8pt}c@{\hskip 3.8pt} @{\hskip 3.8pt}c@{\hskip 3.8pt} @{\hskip 3.8pt}c@{\hskip 3.8pt}}
        I &&&&&&& \\
        &I &&&&&& \\
        &&I&&&&& \\
        &&&I&&&& \\
        &&&&I&&& \\
        &&&&&I&& \\
        &&&&&&I& \\
        &&&&&&&\tilde{\Omega}^2 \\
    \end{array}
    \end{pmatrix}\begin{pmatrix}
    \begin{array}{@{\hskip 3.6pt}c@{\hskip 3.6pt} @{\hskip 3.6pt}c@{\hskip 3.6pt} @{\hskip 3.6pt}c@{\hskip 3.6pt} @{\hskip 3.6pt}c@{\hskip 3.6pt} @{\hskip 3.6pt}c@{\hskip 3.6pt} @{\hskip 3.6pt}c@{\hskip 3.6pt} @{\hskip 3.6pt}c@{\hskip 3.6pt} @{\hskip 3.6pt}c@{\hskip 3.6pt}}
        I&&&&&&& \\
        &I&&&&&& \\
        &&I&&&&& \\
        &&&I&&&& \\
        &&&&I&&& \\
        &&&&&I&& \\
        &&&&&&\tilde{\Omega}^\dagger& \\
        &&&&&&&\tilde{\Omega}^\dagger \\
    \end{array}
    \end{pmatrix}\begin{pmatrix}
    \begin{array}{@{\hskip 3.8pt}c@{\hskip 3.8pt} @{\hskip 3.8pt}c@{\hskip 3.8pt} @{\hskip 3.8pt}c@{\hskip 3.8pt} @{\hskip 3.8pt}c@{\hskip 3.8pt} @{\hskip 3.8pt}c@{\hskip 3.8pt} @{\hskip 3.8pt}c@{\hskip 3.8pt} @{\hskip 3.8pt}c@{\hskip 3.8pt} @{\hskip 3.8pt}c@{\hskip 3.8pt}}
        I&&&&&&& \\
        &I&&&&&& \\
        &&I&&&&& \\
        &&&I&&&& \\
        &&&&I&&& \\
        &&&&&I&& \\
        &&&&&&I& \\
        &&&&&&&\tilde{\Omega}^{\dagger 4} \\
    \end{array}
    \end{pmatrix}$}.
\end{gathered}
\end{equation}

\subsubsection{Circuit implementation}
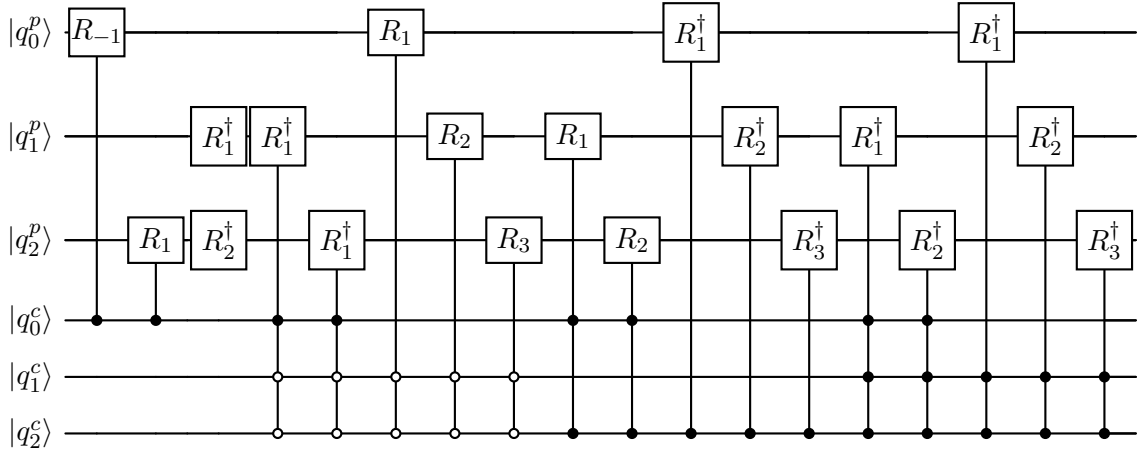
\begin{figure}[h]
\centering
\begin{tikzpicture}
    \node (circ) {
    \begin{quantikz}[row sep=0.6cm, column sep=0.05cm]
    \lstick{$\ket{q_0^p}$} & \gate{\makebox[0.45cm]{$R_{-1}$}} & \qw & \qw & \qw & \qw & \qw & \gate{\makebox[0.45cm]{$R_{1}$}} & \qw & \qw & \qw & \qw & \gate{\makebox[0.45cm]{$R_{1}^\dagger$}} & \qw & \qw & \qw & \qw & \gate{\makebox[0.45cm]{$R_{1}^\dagger$}} & \qw & \qw & \qw \\
    \lstick{$\ket{q_{1}^p}$} & \qw & \qw & \qw & \gate{\makebox[0.45cm]{$R_{1}^\dagger$}} & \gate{\makebox[0.45cm]{$R_{1}^\dagger$}} & \qw & \qw & \gate{\makebox[0.45cm]{$R_{2}$}} & \qw & \gate{\makebox[0.45cm]{$R_{1}$}} & \qw & \qw & \gate{\makebox[0.45cm]{$R_{2}^\dagger$}} & \qw & \gate{\makebox[0.45cm]{$R_{1}^\dagger$}} & \qw & \qw & \gate{\makebox[0.45cm]{$R_{2}^\dagger$}} & \qw & \qw \\
    \lstick{$\ket{q_{2}^p}$} & \qw & \gate{\makebox[0.45cm]{$R_{1}$}} & \qw & \gate{\makebox[0.45cm]{$R_{2}^\dagger$}} & \qw & \gate{\makebox[0.45cm]{$R_{1}^\dagger$}} & \qw & \qw & \gate{\makebox[0.45cm]{$R_{3}$}} & \qw & \gate{\makebox[0.45cm]{$R_{2}$}} & \qw & \qw &  \gate{\makebox[0.45cm]{$R_{3}^\dagger$}} & \qw & \gate{\makebox[0.45cm]{$R_{2}^\dagger$}} & \qw & \qw & \gate{\makebox[0.45cm]{$R_{3}^\dagger$}} & \qw \\
    \lstick{$\ket{q_{0}^c}$} & \ctrl{-3} & \ctrl{-1} & & & \ctrl{-2} & \ctrl{-1} & & & & \ctrl{-2} & \ctrl{-1} & & & & \ctrl{-2} & \ctrl{-1} & & & & \qw \\
    \lstick{$\ket{q_{1}^c}$} & & & & & \octrl{-1} & \octrl{-1} & \octrl{-4} & \octrl{-3} & \octrl{-2} & & & & & & \ctrl{-1} & \ctrl{-1} & \ctrl{-4} & \ctrl{-3} & \ctrl{-2} & \qw \\
    \lstick{$\ket{q_{2}^c}$} & & & & & \octrl{-1} & \octrl{-1} & \octrl{-1} & \octrl{-1} & \octrl{-1} & \ctrl{-2} & \ctrl{-2} & \ctrl{-5} & \ctrl{-4} & \ctrl{-3} & \ctrl{-1} & \ctrl{-1} & \ctrl{-1} & \ctrl{-1} & \ctrl{-1} & \qw \\
    \end{quantikz}
    };
\end{tikzpicture}
\caption{Optimized quantum circuit implementation of the phase shift operator $\Sigma$ for the DTQW on the 8-Cayley graph with generating set $\mathcal{S}=\{\pm1, \pm2, \pm3, 4\}$. The circuit illustrates the sequence of controlled rotation gates acting on the position register $\ket{q^p}$ conditioned on the coin register $\ket{q^c}$, following the three-stage decomposition.}
\label{fig:8_cayley_circuit}
\end{figure}

We now evaluate the gate requirements for each factor of the decomposed phase shift operator. The following analysis details the control logic and gate counts necessitated to implement each factor, as illustrated in the quantum circuit shown in Fig.~\ref{fig:8_cayley_circuit}.

\paragraph{Matrix $A_0^{2}$}
This component implements the diagonal phase block $\tilde{\Omega}^{4}$ conditioned on the set of coin states
$\{\ket{001},\, \ket{011},\, \ket{101},\, \ket{111}\}$. These states are uniquely identified by the least significant coin qubit being in state $\ket{1}$ ($\ket{q_0^c} = \ket{1}$). Since the rotation is independent of the first two coin qubits, it is implemented efficiently using two $C^{(1)}R_{\ell-1}$ gates. 

\paragraph{Matrix $D_0^2$}
The matrix $D_0^2$ applies the inverse diagonal phase blocks $\tilde{\Omega}^{\dagger 2}$ across all coin states. As this operation is entirely coin-independent, it removes the need for control logic, requiring only two single-qubit gates $R^\dagger_\ell$ gates.

\paragraph{Matrix $\mathbf{A}_1$}
$\mathbf{A}_1$ implements $\tilde{\Omega}^{\dagger2}$ conditioned exclusively on the specific coin state $\ket{001}$. Targeting a single basis state requires a higher degree of control, specifically two $C^{(3)}R^\dagger_{\ell}$ gates.

\paragraph{Matrix $\mathbf{D}_1$}
The $\tilde{\Omega}$ blocks in $\mathbf{D}_1$ conditioned on the first two coin states, $\ket{000}$ and $\ket{001}$, where the two most significant qubits are in state $\ket{00}$ ($\ket{q_2^cq_1^c} = \ket{00}$). Consequently, this operation is implemented using three $C^{(2)}R_{\ell+1}$ gates.

\paragraph{Matrix $A_2$}
The implementation of $\tilde{\Omega}^2$ blocks depends on coin states $\ket{101}$ and $\ket{111}$. These states are identified by the first and last qubits being in state $\ket{1}$ ($\ket{q_2^cq_0^c} = \ket{11}$), requiring two $C^{(2)}R_{\ell}$ gates. 

\paragraph{Matrix $D_2$}
The $\tilde{\Omega}^\dagger$ rotations in $D_2$ are conditioned on the four coin states \linebreak $\{\ket{100}, \ket{101}, \ket{110}, \ket{111}\}$, defined by the most significant coin qubit being state $\ket{1}$ ($\ket{q_2^c}=\ket{1}$). This requires three $C^{(1)}R^\dagger_{\ell+1}$ gates. 

\paragraph{Matrix $A_3$}
$A_3$ is merged with the involution block $\mathcal{I}_{\text{inv}}$, resulting in an inverse diagonal phase block $\tilde{\Omega}^{\dagger2}$ conditioned on the final coin state $\ket{111}$. Since it requires all three coin qubits to uniquely identify the state, the implementation requires two $C^{(3)}R^\dagger_{\ell}$ gates.

\paragraph{Matrix $D_3$}
Finally, $D_3$ implements $\tilde{\Omega}^\dagger$ blocks conditioned on the coin states $\ket{110}$ and $\ket{111}$. These states are identified by the first two coin qubits being in state $\ket{11}$ ($\ket{q_2^cq_1^c} = \ket{11}$), requiring three $C^{(2)}R^\dagger_{\ell+1}$ gates to implement. 

\subsubsection{Gate Complexity and Upper Bound CNOT Cost}
\label{sec:upper_bound_CNOT_cost_1D}
The advantages of the proposed three-stage decomposition are summarized in Table~\ref{tab:gate_complexity_1D}, which compares the controlled gate complexity and upper-bound CNOT cost for the phase shift operator. By applying the decomposition, we effectively shift the implementation burden from high-degree controlled operations to a more manageable set of low-degree and single-qubit operations. 

Although the total number of individual controlled gates increases following the decomposition, this trade-off is still beneficial for practical implementation within specific regimes of $k$, as demonstrated by the benchmarking results in Section~\ref{sec:1D_results}. In NISQ-era architectures, multi-qubit controlled gates are decomposed into a sequence of elementary single- and two-qubit gates (CNOT). Extensive research has focused on optimizing the decomposition $n$-controlled $U(2)$ and $SU(2)$ gates to minimize the CNOT cost, where $n$ denotes the number of control qubits. 

To the best of our knowledge, the method proposed by Rosa et al. \cite{rosa_optimizing_2025} provides the most efficient decomposition for the $C^{(\alpha)}R_\lambda$ gates used in this work. Following their approach, an $n$-controlled $U(2)$ operation is first separated into its equivalent $SU(2)$ component and a local phase rotation. The entire gate is then implemented using the optimized scaling provided by Vale et al. \cite{vale_decomposition_nodate} for $SU(2)$ operators, which yields an upper-bound CNOT cost of $16n-24$, and a single auxiliary qubit to correct the local phase. The $1_0$ in Table~\ref{tab:gate_complexity_1D} denotes this auxiliary qubit initialized in the $\ket{0}$ state. The auxiliary resource can be reused multiple times throughout the circuit, effectively keeping its resource cost constant $O(1)$. In the absence of an auxiliary qubit, the CNOT cost would instead follow the quadratic $O(n^2)$ scaling \cite{rosa_optimizing_2025, zindorf_efficient_2025}. We consistently apply the linear $16n-24$ cost model across our analysis to provide a benchmark for the total CNOT requirements of each implementation.

The comparison presented in Table~\ref{tab:gate_complexity_1D} demonstrate a reduction in total CNOT complexity almost by half, from 720 to 362. This improvement occurs despite the increase in total controlled gate count due to the distribution of high-degree controlled gates to lower-degree gates. While the decomposition method allows for $O(16n)$ linear scaling, the high constant factor associated with this remains costly. A single high-degree controlled gate still imposes a much larger CNOT overhead than a collection of lower-degree gates. By minimizing the control degree $\alpha$ across the circuit, we reduce the cumulative resource requirement and improve the overall fidelity of the operation on hardware.

\begin{table}[h]
\centering
\begin{tabular}{lcccc|c}
\hline
Implementation 
& $C^{(3)}R$ 
& $C^{(2)}R$ 
& $C^{(1)}R$ 
& Single-qubit $R$
& CNOT (upper bound)\\
\hline
Naive 
& 18
& 0 
& 0 
& 0 
& $720/1_0$ \\
Decomposed
& 4
& 8
& 5
& 2
& $362/1_0$\\
\hline
\end{tabular}
\caption{Gate complexity comparison for implementation of $\Sigma$ for the 8-Cayley graph with generating set $\mathcal{S}=\{\pm1, \pm2, \pm3, 4\}$ before and after the decomposition.}
\label{tab:gate_complexity_1D}
\end{table}

\subsection{CNOT Complexity Scaling Analysis for 1D Cayley Graphs}
\label{sec:1D_results}

\begin{figure}
    \centering
    \includegraphics[width=1\linewidth]{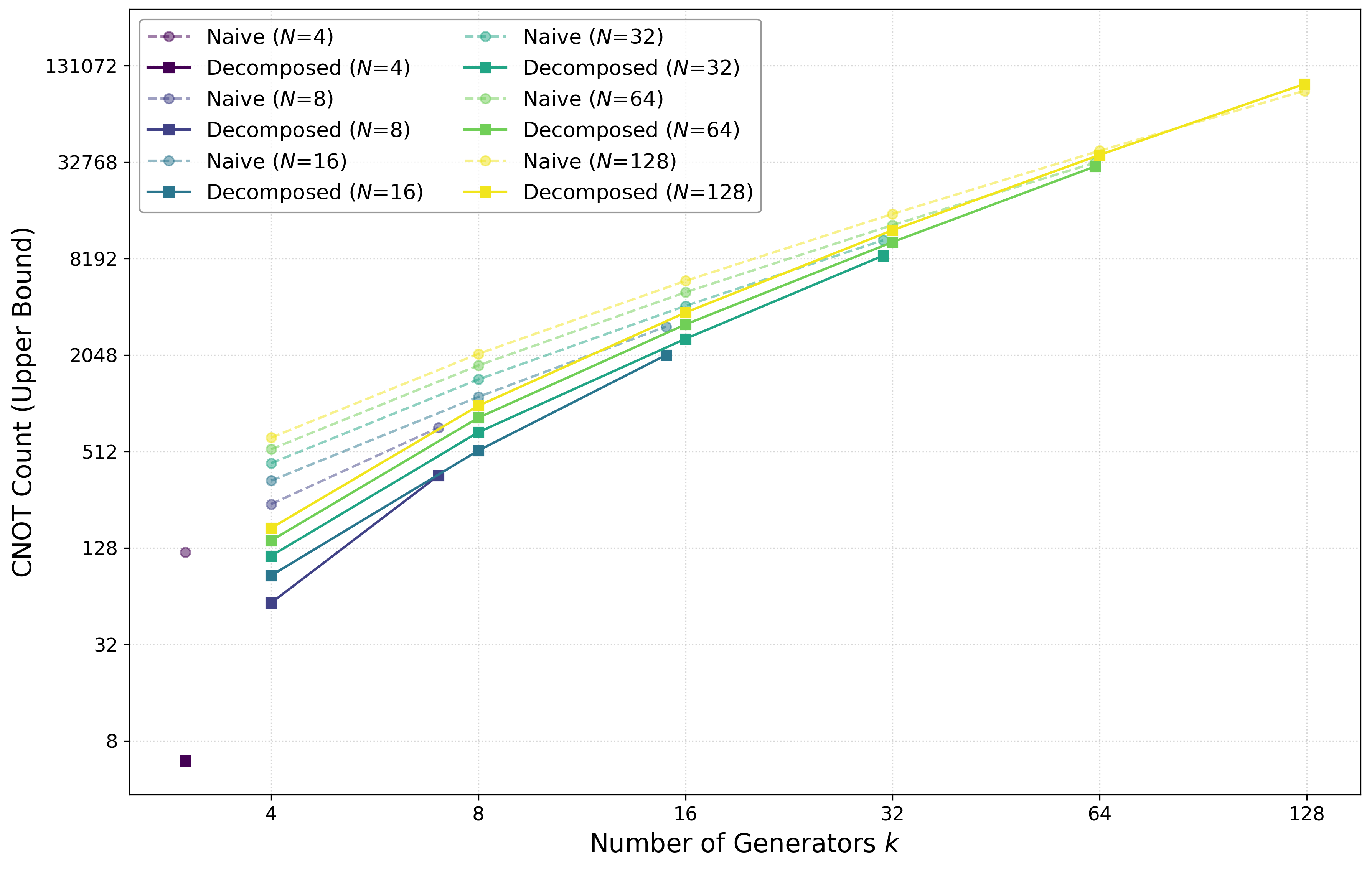}
    \caption{Upper-bound CNOT count as a function of the generator degree $k$ for the naive (dashed) and decomposed (solid) implementations of the phase shift operator $\Sigma$ on 1D inverse-closed Cayley graphs, shown for system sizes $N \in \{4, 8, 16, 32, 64, 128\}$. Both axes are logarithmic. The decomposed implementation consistently achieves a lower CNOT count for $k \leq 64$, beyond which the accumulation of lower-degree controlled gates begins to outweigh the benefit of reduced control degree.}
    \label{fig:IC_cnot_complexity_plot_k_log}
\end{figure}

\begin{figure}
    \centering
    \includegraphics[width=1\linewidth]{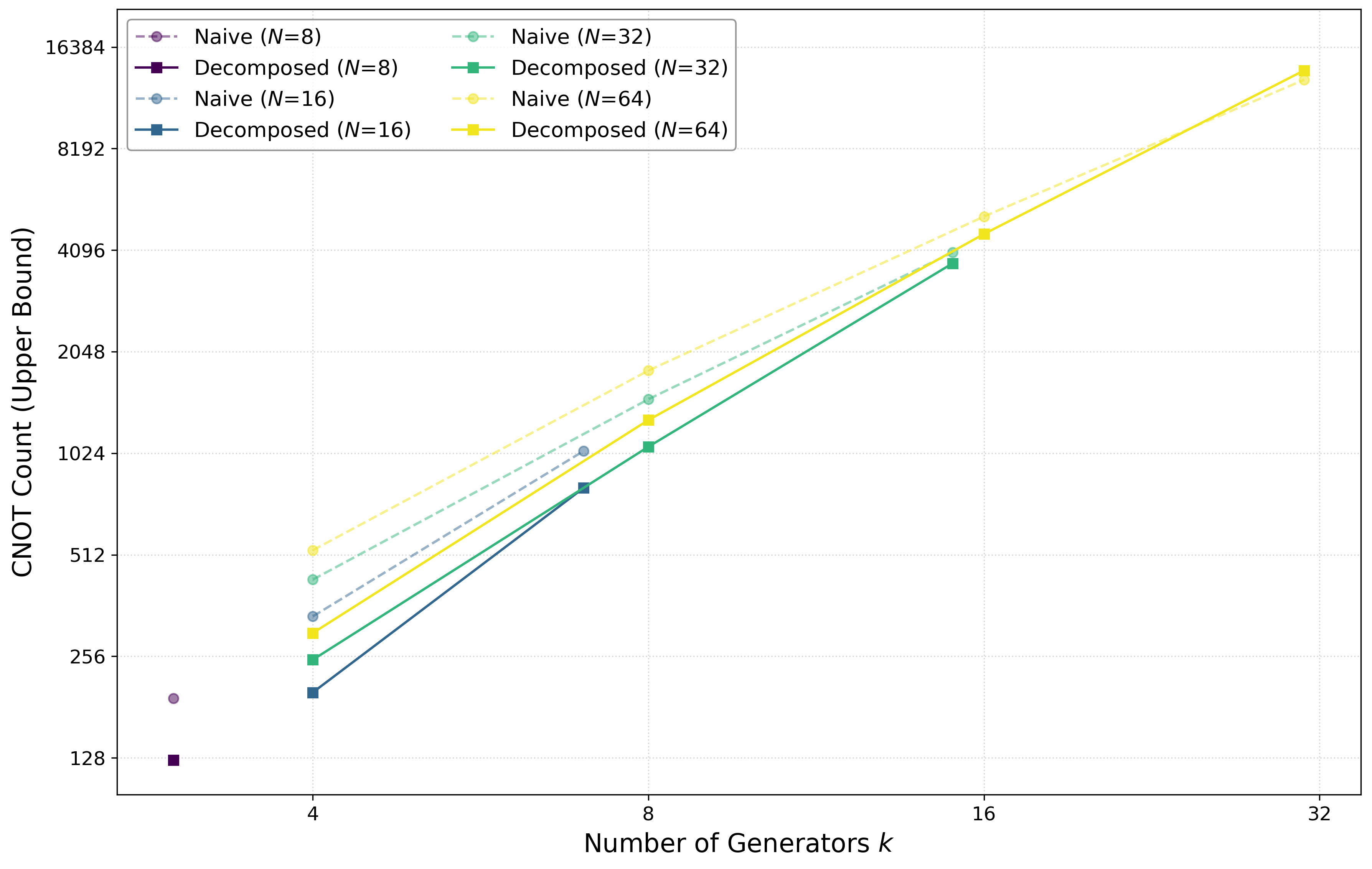}
    \caption{ Upper-bound CNOT count as a function of the generator degree $k$ for the naive (dashed) and decomposed (solid) implementations on 1D non-inverse-closed Cayley graphs, shown for system sizes $N \in \{8, 16, 32, 64\}$. Both axes are logarithmic. The decomposed method remains advantageous for $k \leq 16$.}
    \label{fig:NIC_cnot_complexity_plot_k_log}
\end{figure}

\begin{figure}
    \centering
    \includegraphics[width=1\linewidth]{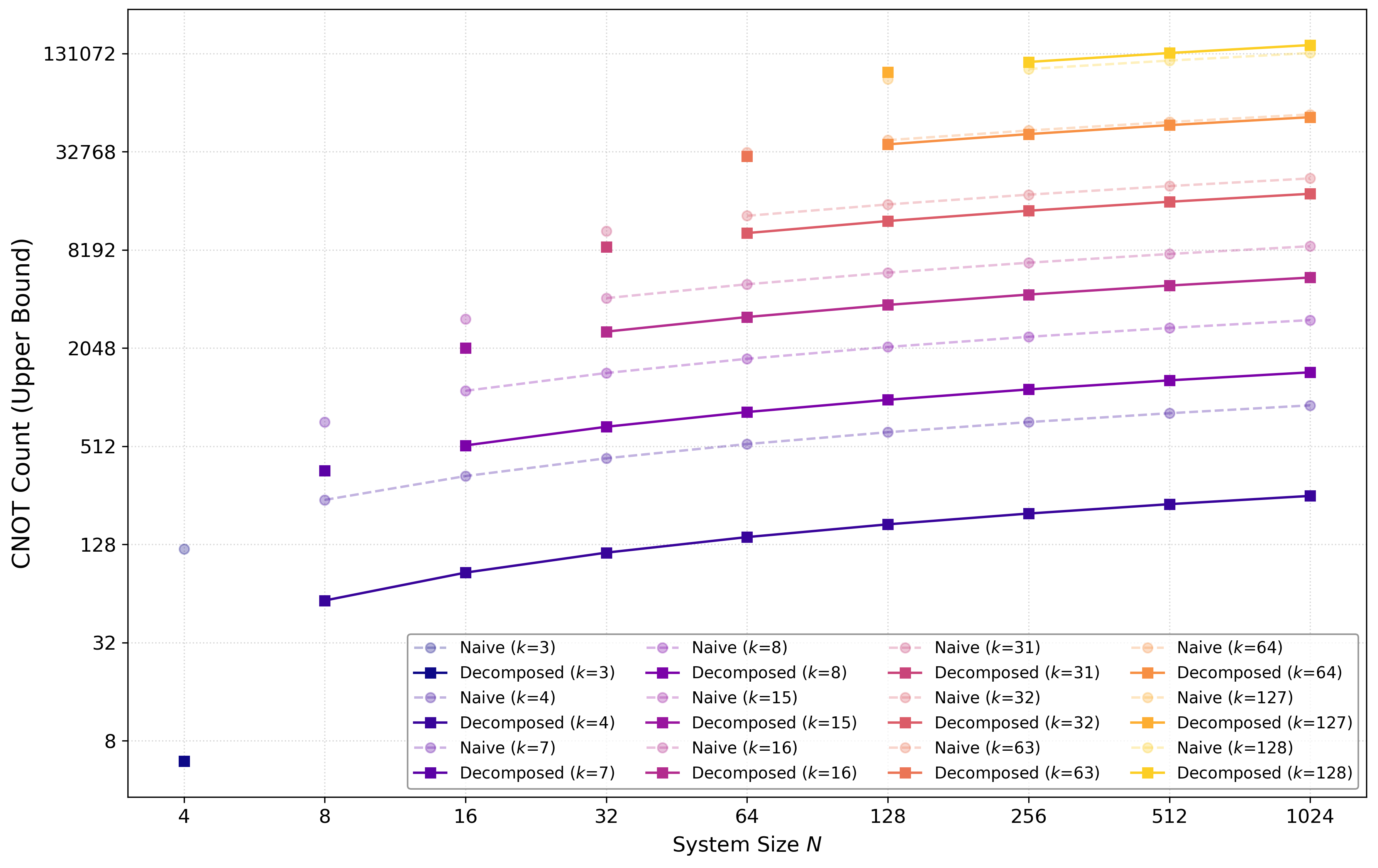}
    \caption{Upper-bound CNOT count as a function of system size $N$ for the naive (dashed) and decomposed (solid) implementations on 1D inverse-closed Cayley graphs, shown for representative values of $k$. Both axes are logarithmic. The two implementations scale similarly with $N$, with a near-constant vertical separation for each fixed $k$.}
    \label{fig:IC_cnot_complexity_plot_N_log}
\end{figure}

\begin{figure}
    \centering
    \includegraphics[width=1\linewidth]{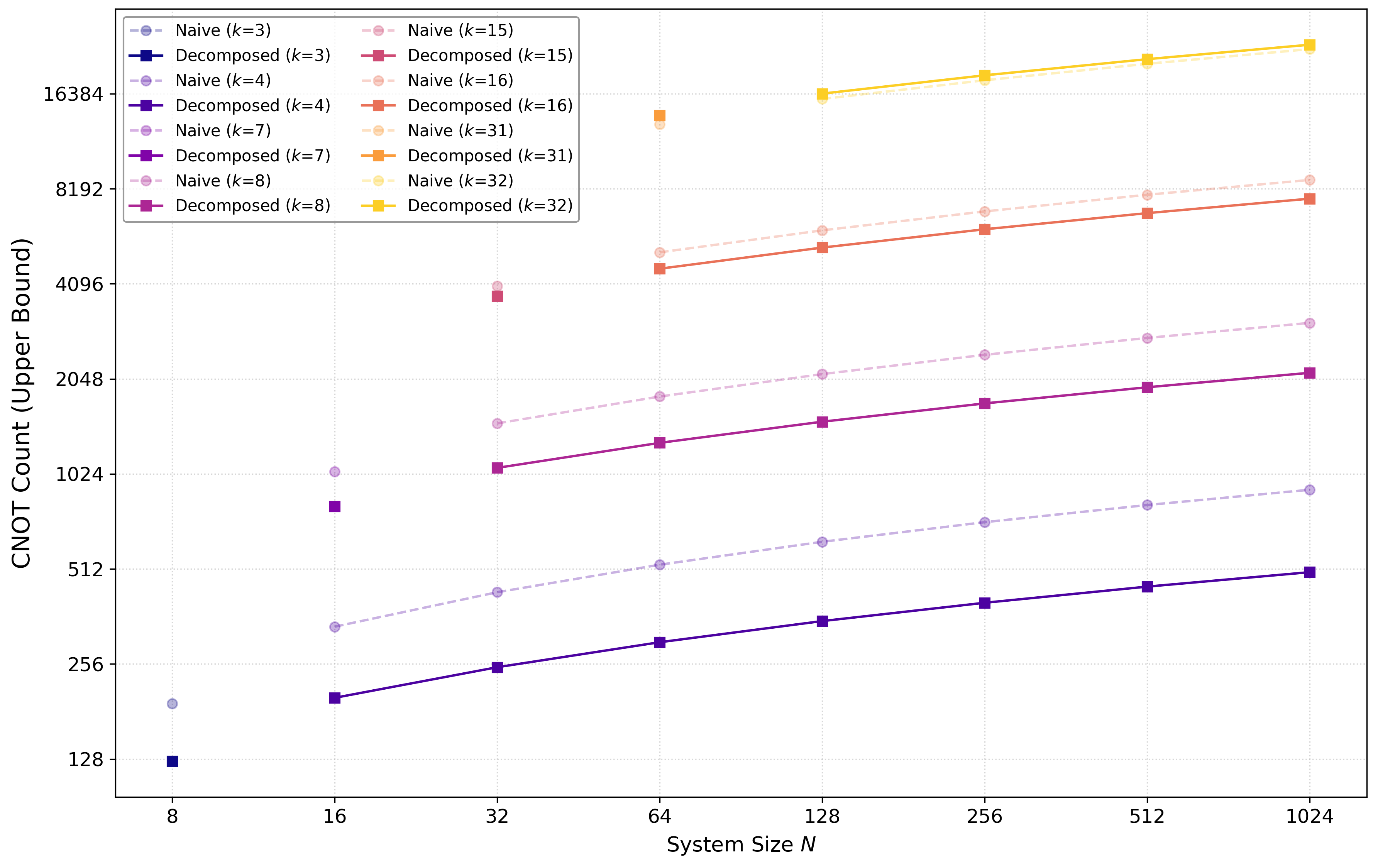}
    \caption{Upper-bound CNOT count as a function of system size $N$ for the naive (dashed) and decomposed (solid) implementations on 1D non-inverse-closed Cayley graphs, shown for representative values of $k$. Both axes are logarithmic. As in the inverse-closed case, the scaling with $N$ is uniform across both methods, and the relative advantage of the decomposed implementation for small $k$ persists independent of system size.}
    \label{fig:NIC_cnot_complexity_plot_N_log}
\end{figure}

To evaluate the relative efficiency of the naive and decomposed implementations, we benchmark the upper-bound CNOT counts as functions of the generator degree $k$ and the system size $N$. Results for both the inverse-closed and non-inverse-closed cases are presented in Figs.~\ref{fig:IC_cnot_complexity_plot_k_log}-\ref{fig:NIC_cnot_complexity_plot_N_log}. 

The benchmarking reveals a critical performance threshold in both cases. In the log-log plots of Figs.~\ref{fig:IC_cnot_complexity_plot_k_log} and ~\ref{fig:NIC_cnot_complexity_plot_k_log}, the naive implementation exhibits near-linear scaling, consistent with a power-law growth in $k$. By contrast, the decomposed implementation displays a sub-linear curvature in log-log space, reflecting the redistribution of the control logic across a larger number of lower-degree gates achieved by the hierarchical factorization. Consequently, the decomposed method yields a strictly lower CNOT count within the regime $k \leq 64$.

Figs.~\ref{fig:IC_cnot_complexity_plot_N_log} and ~\ref{fig:NIC_cnot_complexity_plot_N_log}
 complement this picture by fixing $k$ and varying $N$. The gap between the two methods remains approximately constant as $N$ grows, indicating that the system size does not govern the relative efficiency of the two approaches, and the generator degree $k$ is the dominant parameter. For the inverse-closed case, the decomposed method incurs a higher CNOT cost than the naive implementation for $k>64$; the analogous crossover occurs earlier, at $k>16$, for the non-inverse-closed case.

This crossover behavior is expected due to the linear CNOT scaling in the control degree $\alpha \sim \log_2k$, as established in Section~\ref{sec:upper_bound_CNOT_cost_1D}. The hierarchical decomposition reduces $\alpha$ per gate at the cost of increasing the total number of controlled gates. Because the CNOT overhead scales as $O(16\alpha)$ per gate, lowering $\alpha$ is initially beneficial, however, as $k$ grows, the proliferation of lower-degree gates eventually dominates, and the cumulative cost exceeds that of the naive, high-degree implementation. The threshold value of $k$ therefore marks the point at which the constant prefactors of these two competing contributions intersect. 

\section{DTQW on the $N\times M$ Cayley Graph}
\label{2D_cayley_graph}
A 2D $N \times M$ Cayley graph can be realized as a torus grid graph formed by the Cartesian product of cyclic groups $\mathbb{Z}_N$ and $\mathbb{Z}_M$. This graph is defined as $\Gamma(\mathbb{Z}_N \times \mathbb{Z}_M, \mathcal{S})$, where generating set (assuming no involution) is $\mathcal{S}=\{(\pm1,0),\dots, (\pm k_N/2, 0), (0,\pm1),\dots,  (0,\pm k_M/2)\}$. Here, $k_N$ and $k_M$ denote the degrees associated with the periodic dimension $\mathbb{Z}_N$ and $\mathbb{Z}_M$, respectively. The setup follows the 1D architecture introduced in Sec.~\ref{sec:circuit_setup}, with $K=k_N+k_M$ being the new coin space dimension and the position space now possessing two degrees of freedom. The associated Hilbert spaces are $\mathcal{H}_p^{(N)}=\text{span}(\{\ket{x}: x = 0,1,\dotsm N-1\})$ and $\mathcal{H}_p^{(M)}=\text{span}(\{\ket{y_i}: y=0,1,\dots, M-1\})$. The full Hilbert space of the walk is defined as:
\begin{equation}
    \mathcal{H} = \text{span}\left\{\ket{c}\ket{x}\ket{y}: c=0,1,\dots , K - 1; x = 0,1,\dots, N-1; y=0,1,\dots, M-1\}\right),
\end{equation}
where the first $k_N$ coin states govern position shifts in $\mathbb{Z}_N$ and the subsequent $k_M$ coin states govern position shifts in $\mathbb{Z}_M$. The state of the walker is encoded using $\alpha= \lceil \log_2 K\rceil$ coin qubits, $n = \log_2N$ position qubits for $x$, and $m = \log_2 M$ position qubits for $y$:
\begin{equation}
    \ket{\psi}_{walker} = \ket{c}\ket{x}\ket{y}=\ket{q_\alpha^cq_{\alpha-1}^c\cdots q_0^c}\ket{q_n^p q_{n-1}^p\cdots q_0^p}\ket{q_m^p q_{m-1}^p\cdots q_0^p}.
\end{equation}

\subsection*{Shift Operator and Diagonalization}
The shift operator $S$ is a block diagonal matrix of size $KNM \times KNM$ composed of two primary blocks, $S_{\mathbb{Z}_N}$ and $S_{\mathbb{Z}_M}$: 
\begin{equation}
\begin{gathered}
    S = \begin{pmatrix}
        S_{\mathbb{Z}_N} & \\
        & {\mathbb{Z}_M}
    \end{pmatrix}, \quad \text{where} \\[6pt]
S_{\mathbb{Z}_N} =            \begin{pmatrix}
    \begin{array}{@{\hskip -.5pt}c@{\hskip -12pt}c@{\hskip -12pt}c@{\hskip -12pt}c@{\hskip -13pt}c@{\hskip -.5pt}}
        P_0\otimes I_M &&&& \\
        & P_0^\mathsf{T}\otimes I_M &&& \\
        && \ddots && \\
        &&& P_0^{k_N/2}\otimes I_M & \\
        &&&& P_0^{\mathsf{T}k_N/2}\otimes I_M \\
    \end{array}
    \end{pmatrix},
    \Sigma_M =
    \begin{pmatrix}
    \begin{array}{@{\hskip -.5pt}c@{\hskip -12pt}c@{\hskip -12pt}c@{\hskip -12pt}c@{\hskip -13pt}c@{\hskip -.5pt}}
        I_N\otimes P_0 &&&& \\
        & I_N\otimes P_0^\mathsf{T} &&& \\
        && \ddots && \\
        &&& I_N\otimes P_0^{k_M/2} & \\
        &&&& I_N\otimes P_0^{\mathsf{T}k_M/2} \\
    \end{array}
    \end{pmatrix}
\end{gathered}
\end{equation}
where $S_{\mathbb{Z}_N}$ has dimensions $k_NNM\times k_NNM$ and $S_{\mathbb{Z}_M}$ has dimensions $k_MNM\times k_MNM$. 

$S$ is diagonalized using the SWAP-free QFT and IQFT matrices. The composite QFT operator $\tilde{\mathcal{F}}$ is defined as:
\begin{equation}
\begin{gathered}
    \tilde{\mathcal{F}} = \begin{pmatrix}
        \tilde{\mathcal{F}}_{\mathbb{Z}_N} & \\
        & \tilde{\mathcal{F}}_{\mathbb{Z}_M}
\end{pmatrix},\quad \text{where} \\[6pt]
    \tilde{\mathcal{F}}_{\mathbb{Z}_N} =
    \left.
    \begin{pmatrix}
    \begin{array}{@{\hskip 0pt}c@{\hskip -3pt}c@{\hskip -3pt}c@{\hskip -3pt}c@{\hskip 0pt}}
        \tilde{\mathcal{F}}_N\otimes I_M &&&\\
        & \tilde{\mathcal{F}}_N \otimes I_M &&\\
        &&\ddots&\\
        &&& \tilde{\mathcal{F}}_N \otimes I_M\\
    \end{array}
    \end{pmatrix}
    \right\}
    k_N, \quad \tilde{\mathcal{F}}_{\mathbb{Z}_M} = 
    \left.\begin{pmatrix}
    \begin{array}{@{\hskip 0pt}c@{\hskip -3pt}c@{\hskip -3pt}c@{\hskip -3pt}c@{\hskip 0pt}}
        I_N \otimes \tilde{\mathcal{F}}_M &&&\\
        & I_N \otimes \tilde{\mathcal{F}}_M &&\\
        && \ddots &\\
        &&& I_N \otimes \tilde{\mathcal{F}}_M \\
    \end{array}
    \end{pmatrix}
    \right\}k_M,
\end{gathered}
\end{equation}

The resulting phase shift operator $\Sigma$ is given by:
\begin{equation}
\begin{gathered}
\Sigma =
    \begin{pmatrix}
        \Sigma_{\mathbb{Z}_N} & \\
        & \Sigma_{\mathbb{Z}_M}
    \end{pmatrix},
    \quad\text{where} \\[6pt]
    \Sigma_{\mathbb{Z}_N} =
    \begin{pmatrix}
    \begin{array}{@{\hskip -.5pt}c@{\hskip -9.5pt}c@{\hskip -9.5pt}c@{\hskip -9.5pt}c@{\hskip -9.5pt}c@{\hskip -.5pt}}
        \tilde{\Omega}^\dagger\otimes I_M &&&& \\
        & \tilde{\Omega}\otimes I_M &&& \\
        && \ddots && \\
        &&& \tilde{\Omega}^{\dagger k_N/2}\otimes I_M & \\
        &&&& \tilde{\Omega}^{k_N/2}\otimes I_M \\
    \end{array}
    \end{pmatrix},
    \Sigma_{\mathbb{Z}_M} =
    \begin{pmatrix}
    \begin{array}{@{\hskip -.5pt}c@{\hskip -9.5pt}c@{\hskip -9.5pt}c@{\hskip -9.5pt}c@{\hskip -9.5pt}c@{\hskip -.5pt}}
        I_N\otimes \tilde{\Omega}^\dagger &&&& \\
        & I_N\otimes \tilde{\Omega} &&& \\
        && \ddots && \\
        &&& I_N\otimes \tilde{\Omega}^{\dagger k_M/2} & \\
        &&&& I_N\otimes \tilde{\Omega}^{k_M/2} \\
    \end{array}
    \end{pmatrix}
\end{gathered}
\end{equation}

\subsection*{Decomposition and Implementation}
The same decomposition method from the 1D case can be applied for each $\Sigma_{\mathbb{Z}_N}$ and $\Sigma_{\mathbb{Z}_M}$. The gate complexity analysis for the implementation of each $\Sigma_{\mathbb{Z}_N}$ and $\Sigma_{\mathbb{Z}_M}$ is exactly the same as described in Sec.~\ref{sec:gate_complexity}, however, the decomposed operators within $\Sigma_{\mathbb{Z}_N}$ act exclusively on the position register $\mathcal{H}^{(N)}_p$, while $\Sigma_{\mathbb{Z}_M}$ acts only $\mathcal{H}_p^{(M)}$. This separation enforces a modular circuit design where the two dimensions are updated conditionally based on the state of the coin register. A concrete implementation of this procedure on a specific 2D Cayley graph is provided in Sec.~\ref{sec:2D torus graph example}.

\subsection{Extension to a General $d$-Dimensional Cayley graph}
The implementation of the DTQW on a 2D $N\times M$ Cayley graph can be generalized to an arbitrary $d$-dimensional torus defined by the group $\mathbb{Z}_{N_1} \times \mathbb{Z}_{N_2} \times \cdots \times  \mathbb{Z}_{N_d}$. Extending the walk to higher dimensions directly translates to adding position Hilbert spaces of size $N_1, N_2, \dots ,N_d$ and their respective position registers. The full Hilbert space of the walk is then characterized as:
\begin{equation}
    \mathcal{H}=\mathcal{H}_c^{(K)}\otimes \mathcal{H}_p^{(N_1)} \otimes \mathcal{H}_p^{(N_2)} \otimes \cdots \otimes \mathcal{H}_p^{(N_d)},
\end{equation}
where the coin space dimension is $ K = \sum_{i=1}^d  k_{N_i}$, representing the sum of the degrees of the generating sets associated with each periodic dimension. 

The QFT and IQFT matrices used to diagonalize the shift operator are extended to the $d$-dimensional case accordingly. The composite QFT operator is defined by the block diagonal structure $\tilde{\mathcal{F}}=\text{diag}\left(\tilde{\mathcal{F}}_{\mathbb{Z}_{N_1}}, \tilde{\mathcal{F}}_{\mathbb{Z}_{N_2}}, \dots, \tilde{\mathcal{F}}_{\mathbb{Z}_{N_d}} \right)$, where each $\tilde{\mathcal{F}}_{\mathbb{Z}_{N_i}}$ corresponds to the $i$-th dimension and consists of $k_{N_i}$ identical blocks. Each of these blocks is defined by the tensor product:
\begin{equation}
     B_1 \otimes B_2 \otimes \cdots \otimes B_d, \quad \text{where} \quad B_j = \begin{cases} \tilde{\mathcal{F}}_{N_i} & j = i \\ I_{N_j} & j \neq i \end{cases}
\end{equation}
This ensures that for a given dimension index $i$, the QFT acts only on the corresponding $i$-th position register while acting as the identity on all others.

The resulting phase shift operator $\Sigma$ is expressed as:
\begin{equation}
    \Sigma =
    \begin{pmatrix}
        \Sigma_{N_1} & & & & &\\
        & \Sigma_{N_2} & & & &\\
        & & \ddots & & &\\
        & & & \Sigma_{N_i} & &\\
        & & & & \ddots &\\
        & & & & &\Sigma_{N_d}.
    \end{pmatrix}
\end{equation}
Each block $\Sigma_{N_i}$ corresponds to the phase shifts for the $i$-th dimension and is defined as:
\begin{equation}
    \begin{gathered}
    \Sigma_{N_i} =
    \begin{pmatrix}
    \tilde{\Omega}^\dagger_{(i)} &&&& \\
    & \tilde{\Omega}_{(i)} &&& \\
    && \ddots && \\
    &&& \tilde{\Omega}^{\dagger k_{N_i}/2}_{(i)} & \\
    &&&& \tilde{\Omega}^{k_{N_i}/2}_{(i)}
    \end{pmatrix}.
\end{gathered}
\end{equation}
Following the same logic as the QFT operators, the component blocks $\tilde{\Omega}^\dagger_{(i)}$ are generalized as tensor products acting on the $i$-th position register:  
\begin{equation}
    \tilde{\Omega}^\dagger_{(i)} = B_1 \otimes B_2 \otimes \cdots \otimes B_d, \quad B_{j} = \begin{cases}
        \tilde{\Omega}^\dagger_{N_i} & j=i \\ I_{N_j} & j \neq i
    \end{cases}
\end{equation}

Each $\Sigma_{N_i}$ can be treated as an independent 1D phase shift operator and the decomposition methods established in previous sections (both for cases with and without an involution) remain directly applicable.

\subsection{Example: DTQW on a 2D Torus Graph}
\label{sec:2D torus graph example}

In this section, we present a concrete implementation of the DTQW on a two-dimensional torus grid graph defined by the product group of $\mathbb{Z}_{16} \times \mathbb{Z}_8$. This structure is realized as a 2D Cayley graph $\Gamma(\mathbb{Z}_{16} \times \mathbb{Z}_8, \mathcal{S})$ with degree $K = 8$, where the generating set is $\mathcal{S}=\{(\pm1,0), (\pm2,0),(0,\pm1),(0,\pm2)\}$. The full Hilbert space is defined as:
\begin{equation}
    \mathcal{H} = \text{span}\{\ket{c}\ket{x}\ket{y}: c = 0,\dots,7 ;x=0,\dots,15 ;y = 0,\dots,7\}. 
\end{equation}
The position space requires $n+m  = \log_2 16 + \log_2 8 = 7$ qubits, while the coin space is encoded using $\alpha = \log_2 8 = 3$ qubits.

The shift operator $S$ is a $1024 \times 1024$ block diagonal matrix composed of $S_{\mathbb{Z}_{16}}$ and $S_{\mathbb{Z}_8}$, each of dimension $512 \times 512$. These blocks correspond to the shift operations within their cyclic group factors. 
\begin{equation}
\begin{gathered}
    S = 
    \begin{pmatrix}
        S_{\mathbb{Z}_{16}} & \\
        & S_{\mathbb{Z}_8}
    \end{pmatrix}, \quad \text{where} \\[6pt]
     S_{\mathbb{Z}_{16}} =\begin{pmatrix}
    \begin{array}{@{\hskip 0pt}c@{\hskip -3pt}c@{\hskip -3pt}c@{\hskip -3pt}c@{\hskip 0pt}}
         P_0 \otimes I_8 & & &\\
        & P_0^{\mathsf{T}} \otimes I_8 & & \\
        & & P_0^2 \otimes I_8 & \\
        & & & P_0^{\mathsf{T}2} \otimes I_8
    \end{array}
    \end{pmatrix}, \quad
    S_{\mathbb{Z}_8} =\begin{pmatrix}
    \begin{array}{@{\hskip 0pt}c@{\hskip -3pt}c@{\hskip -3pt}c@{\hskip -3pt}c@{\hskip 0pt}}
         I_{16} \otimes P_0 & & & \\
        & I_{16}\otimes P_0^{\mathsf{T}} & & \\
        & & I_{16} \otimes P_0^2 &\\
        & & & I_{16}\otimes P_0^{\mathsf{T}2}
    \end{array}
    \end{pmatrix}.
\end{gathered}
\end{equation}
Note that the increment operators $P_0$ appearing in $S_{\mathbb{Z}_{16}}$ and $S_{\mathbb{Z}_8}$ have different dimensions, specifically $16\times16$ and $8 \times 8$, respectively. 

Using the composite QFT operator $\tilde{\mathcal{F}} = \text{diag}\left((\tilde{\mathcal{F}}_{16}\otimes I_8)^{\oplus 4},( I_{16} \otimes \tilde{\mathcal{F}}_8)^{\oplus 4}\right)$, the shift operator $S$ is diagonalized into the phase shift operator $\Sigma$: 
\begin{equation}
\begin{gathered}
\Sigma= \tilde{\mathcal{F}}S\tilde{\mathcal{F}}^\dagger=\begin{pmatrix}
        \Sigma_{\mathbb{Z}_{16}} & \\
        & \Sigma_{\mathbb{Z}_8}
    \end{pmatrix},\quad \text{where} \\[6pt]
    \Sigma_{\mathbb{Z}_{16}} = 
    \begin{pmatrix}
    \begin{array}{@{\hskip 0pt}c@{\hskip -3pt}c@{\hskip -3pt}c@{\hskip -3pt}c@{\hskip 0pt}}
        \tilde{\Omega}^\dagger\otimes I_8 & & &\\
        & \tilde{\Omega}\otimes I_8 & & \\
        & & \tilde{\Omega}^{\dagger2}\otimes I_8 \\
        & & & \tilde{\Omega}^2\otimes I_8
    \end{array}
    \end{pmatrix}, \quad \Sigma_{\mathbb{Z}_8} = \begin{pmatrix}
    \begin{array}{@{\hskip 0pt}c@{\hskip -3pt}c@{\hskip -3pt}c@{\hskip -3pt}c@{\hskip 0pt}}
        I_{16} \otimes \tilde{\Omega}^\dagger & & & \\
        & I_{16} \otimes \tilde{\Omega} & & \\
        & & I_{16} \otimes \tilde{\Omega}^{\dagger2} & \\
        & & & I_{16} \otimes \tilde{\Omega}^2
    \end{array}
    \end{pmatrix}.
    \label{eq:68}
\end{gathered}
\end{equation}
Similar to the increment operators, the diagonal phase blocks $\tilde{\Omega}$ in $\Sigma_{\mathbb{Z}_{16}}$ and $\Sigma_{\mathbb{Z}_8}$ differ in dimension, each associated with its respective cyclic group.

\subsubsection{Decomposition Stages}

\paragraph{First-order}
The decomposition is performed for the phase shift operators $\Sigma_{\mathbb{Z}_{16}}$ and $\Sigma_{\mathbb{Z}_8}$. Each is treated as a 1D non-involution case where the degree $k$ is specific to that dimension. By factoring out the block diagonal matrices composed of $\tilde{\Omega}^\dagger$ and $\tilde{\Omega}$, we decompose $\Sigma_{\mathbb{Z}_{16}}$ into $M_0^{\mathbb{Z}_{16}}\cdot M_1^{\mathbb{Z}_{16}}$ and $\Sigma_{\mathbb{Z}_{8}}$ into $M_0^{\mathbb{Z}_{8}}\cdot M_1^{\mathbb{Z}_{8}}$. As a result, the complete phase shift operator $\Sigma$ is expressed as follows:
\begin{equation}
\begin{gathered}
    \Sigma = \text{diag}\left(M_0^{\mathbb{Z}_{16}}, M_0^{\mathbb{Z}_8}\right) \cdot \text{diag}\left(M_1^{\mathbb{Z}_{16}}, M_1^{\mathbb{Z}_8}\right), \quad \text{where} \\[6pt]
    M_0^{\mathbb{Z}_{16}} = 
    \begin{pmatrix}
    \begin{array}{@{\hskip 0pt}c@{\hskip -7pt}c@{\hskip -7pt}c@{\hskip -7pt}c@{\hskip 0pt}}
        \tilde{\Omega}^\dagger\otimes I_8 & & & \\
        & \tilde{\Omega}\otimes I_8 & &\\
        & & \tilde{\Omega}^{\dagger}\otimes I_8 & \\
        & & & \tilde{\Omega}\otimes I_8\\ 
    \end{array}
    \end{pmatrix}, \quad M_0^{\mathbb{Z}_{8}} = \begin{pmatrix}
        \begin{array}{@{\hskip 0pt}c@{\hskip -7pt}c@{\hskip -7pt}c@{\hskip -7pt}c@{\hskip 0pt}}
             I_{16} \otimes \tilde{\Omega}^\dagger & & &\\ 
             & I_{16} \otimes \tilde{\Omega} & &\\ 
             & & I_{16} \otimes \tilde{\Omega}^\dagger &\\ 
             & & & I_{16} \otimes \tilde{\Omega} \\ 
        \end{array}
    \end{pmatrix} \\[6pt]
    M_1^{\mathbb{Z}_{16}} = \begin{pmatrix}
    \begin{array}{@{\hskip 0pt}c@{\hskip -7pt}c@{\hskip -7pt}c@{\hskip -7pt}c@{\hskip 0pt}}
        I_{16} \otimes I_8 & & & \\
        & I_{16}\otimes I_8 & & \\
        & & \tilde{\Omega}^{\dagger}\otimes I_8 &\\
        & & & \tilde{\Omega}\otimes I_8\\ 
    \end{array}
    \end{pmatrix}, \quad M_1^{\mathbb{Z}_{8}} = \begin{pmatrix}
    \begin{array}{@{\hskip 0pt}c@{\hskip -7pt}c@{\hskip -7pt}c@{\hskip -7pt}c@{\hskip 0pt}}
        I_{16} \otimes I_8 & & &\\ 
        & I_{16} \otimes I_8 & &\\ 
        & & I_{16} \otimes \tilde{\Omega}^\dagger &\\ 
        & & & I_{16} \otimes \tilde{\Omega} \\ 
    \end{array}
    \end{pmatrix}.
\end{gathered}
\end{equation}

\paragraph{Second-order}
The second-order decomposition factors the $M_0$ and $M_1$ matrices into products of $A_i$ and $D_i$, which separates coin-state-specific operations from global and local phases. This results in $\Sigma_{\mathbb{Z}_4} =A_0^{\mathbb{Z}_{16}}D_0^{\mathbb{Z}_{16}}A_1^{\mathbb{Z}_{16}}D_1^{\mathbb{Z}_{16}}$ and $\Sigma_{\mathbb{Z}_8}=A_0^{\mathbb{Z}_8}D_0^{\mathbb{Z}_8}A_1^{\mathbb{Z}_8}D_1^{\mathbb{Z}_8}$. Consequently, $\Sigma$ is reformulated as follows:
\begin{equation}
\begin{gathered}
    \Sigma = \text{diag}\left(A_0^{\mathbb{Z}_{16}}, A_0^{\mathbb{Z}_8}\right) \cdot \text{diag}\left(D_0^{\mathbb{Z}_{16}}, D_0^{\mathbb{Z}_8}\right)\cdot \text{diag}\left(A_1^{\mathbb{Z}_{16}}, A_1^{\mathbb{Z}_8}\right) \cdot \text{diag}\left(D_1^{\mathbb{Z}_{16}}, D_1^{\mathbb{Z}_8}\right), \quad \text{where} \\[6pt] 
    A_0^{\mathbb{Z}_{16}} = \left((I_{16}\otimes I_8) \oplus (\tilde{\Omega}^2 \otimes I_8)\right)^{\oplus2}, \quad A_0^{\mathbb{Z}_8} = \left((I_{16}\otimes I_8) \oplus (I_8 \otimes \tilde{\Omega}^2) \right)^{\oplus2}, \\[6pt]
    D_0^{\mathbb{Z}_{16}} = \left(\tilde{\Omega}^\dagger \otimes I_8\right)^{\oplus4}, \quad D_0^{\mathbb{Z}_{8}} = \left(I_{16} \otimes \tilde{\Omega}^\dagger\right)^{\oplus4}, \\[6pt]
    A_1^{\mathbb{Z}_{16}} = (I_{16}\otimes I_8)^{\oplus 3} \oplus (\tilde{\Omega}^2 \otimes I_8), \quad A_1^{\mathbb{Z}_8} = (I_{16}\otimes I_8)^{\oplus 3} \oplus (I_8 \otimes \tilde{\Omega}^2), \\[6pt]
    D_1^{\mathbb{Z}_{16}} = (I_{16} \otimes I_8)^{\oplus2} \oplus (\tilde{\Omega}^\dagger \otimes I_8)^{\oplus2}, \quad D_1^{\mathbb{Z}_{8}} = (I_{16} \otimes I_8)^{\oplus2} \oplus (I_8 \otimes \tilde{\Omega}^\dagger)^{\oplus2}
\end{gathered}
\label{eq:63}
\end{equation}

\paragraph{Third-order}
The third-order decomposition is an optimization stage applicable only when the degree of the generating set for each cyclic group satisfies $k_{N_i}>4$. In this specific 2D torus graph, the degrees are $k_{N_1} = k_{N_2}=4$. Consequently, this stage is not necessary, and the final implementation of the phase shift operator therefore remains in the form established in Eq.~\ref{eq:63}.

\subsubsection{Circuit Implementation}
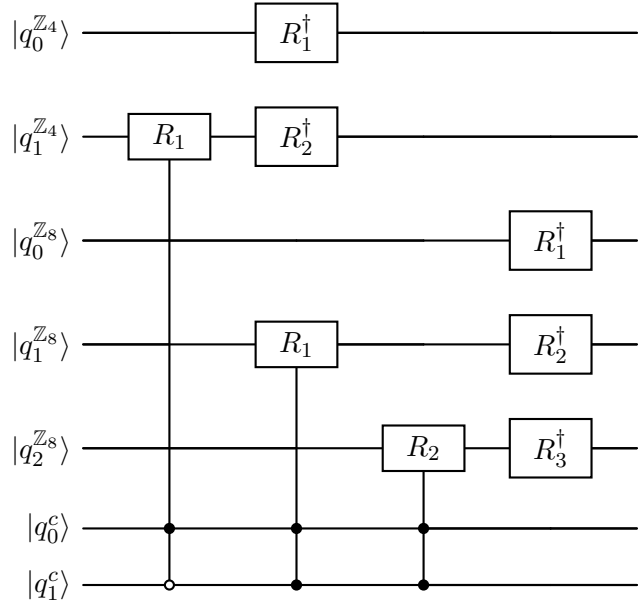
\begin{figure}[h]
\centering
\begin{tikzpicture}
    \node (circ) {
    \begin{quantikz}[row sep=0.6cm, column sep=0.6cm]
    \lstick{$\ket{q_0^{\mathbb{Z}_4}}$} & \qw & \gate{\makebox[0.8cm]{$R_1^\dagger$}} & \qw & \qw & \qw \\
    \lstick{$\ket{q_{1}^{\mathbb{Z}_4}}$}  & \gate{\makebox[0.8cm]{$R_1$}} & \gate{\makebox[0.8cm]{$R_2^\dagger$}} & \qw & \qw & \qw\\
    \lstick{$\ket{q_{0}^{\mathbb{Z}_8}}$} & \qw & \qw & \qw & \gate{\makebox[0.8cm]{$R_1^\dagger$}} &\qw \\
    \lstick{$\ket{q_{1}^{\mathbb{Z}_8}}$} & \qw & \gate{\makebox[0.8cm]{$R_1$}} & \qw & \gate{\makebox[0.8cm]{$R_2^\dagger$}} & \qw \\
    \lstick{$\ket{q_{2}^{\mathbb{Z}_8}}$} & \qw & \qw & \gate{\makebox[0.8cm]{$R_2$}}& \gate{\makebox[0.8cm]{$R_3^\dagger$}} & \qw \\
    \lstick{$\ket{q_{0}^c}$} & \ctrl{-4} & \ctrl{-2} & \ctrl{-1} & \qw & \qw\\
    \lstick{$\ket{q_{1}^c}$} & \octrl{-1} & \ctrl{-1} & \ctrl{-1} & \qw & \qw
    \end{quantikz}
    };
\end{tikzpicture}
\caption{Optimized quantum circuit implementation of the phase shift operator $\Sigma$ for the DTQW on the $\mathbb{Z}_4 \times\mathbb{Z}_8$ Cayley graph with the generating set $\mathcal{S}=\{(\pm1,0),(0,\pm1)\}$. The gates are partitioned to act on the $\mathbb{Z}_4$ position register and the $\mathbb{Z}_8$ position register based on the state of the 2-qubit coin register.}
\label{fig:2d_torus_circuit}
\end{figure}

The circuit is initialized with 7 total position registers (4 for $\mathbb{Z}_{16}$ and 3 for $\mathbb{Z}_8$)and 3 coin registers. The specific gate requirements and control logic for each matrix component are detailed below, following the circuit architecture shown in Fig.~\ref{fig:2d_torus_circuit}.

\paragraph{Matrix $A_0^{\mathbb{Z}_{16}}$} This matrix applies $\tilde{\Omega}^2$ to the $\mathbb{Z}_{16}$ position register, controlled on the coin states $\{\ket{001}, \ket{011}\}$. These states are uniquely identified by the most significant qubit being in state $\ket{0}$ and the least significant qubit being in state $\ket{1}$ ($\ket{q_2^cq_0^c} = \ket{01}$). This implementation requires three $C^{(2)}R_{\ell}$ gates.

\paragraph{Matrix $A_0^{\mathbb{Z}_8}$} $A_0^{\mathbb{Z}_8}$ implements $\tilde{\Omega}^2$ on the $\mathbb{Z}_8$ position register, conditioned on the coin states $\{\ket{101}, \ket{111}\}$. In this case, both the most significant and the least significant qubits are in state $\ket{1}$ ($\ket{q_2^cq_0^c} = \ket{11}$). This operation requires two $C^{(2)}R_{\ell}$ gates. 

\paragraph{Matrix $D_0^{\mathbb{Z}_{16}}$} The matrix $D_0^{\mathbb{Z}_{16}}$ applies $\tilde{\Omega}^\dagger$ to the $\mathbb{Z}_{16}$ position register, conditioned on the most significant qubit being in state $\ket{0}$. Its implementation requires four $C^{(1)}R^\dagger_{\ell+1}$ gates. 

\paragraph{Matrix $D_0^{\mathbb{Z}_8}$} Similarly, $D_0^{\mathbb{Z}_8}$ implements $\tilde{\Omega}^\dagger$ on the $\mathbb{Z}_8$ position register, conditioned on the most significant qubit being in state $\ket{1}$. This requires three $C^{(1)}R^\dagger_{\ell+1}$ gates.  

\paragraph{Matrix $A_1^{\mathbb{Z}_{16}}$} $A_1^{\mathbb{Z}_{16}}$ applies $\tilde{\Omega}^2$ to the $\mathbb{Z}_{16}$ position register, controlled specifically on the coin state $\ket{011}$. To uniquely isolate this state, it requires three $C^{(3)}R_{\ell}$ gates.

\paragraph{Matrix $A_1^{\mathbb{Z}_{8}}$} $A_1^{\mathbb{Z}_{8}}$ implements $\tilde{\Omega}^2$ on the $\mathbb{Z}_{8}$ position register, controlled on the coin state $\ket{111}$. This operation requires two $C^{(3)}R_{\ell}$ gates.

\paragraph{Matrix $D_1^{\mathbb{Z}_{16}}$} The matrix $D_1^{\mathbb{Z}_{16}}$ applies $\tilde{\Omega}^\dagger$ to the $\mathbb{Z}_{16}$ position register, conditioned on the coin states $\{\ket{010}, \ket{011}\}$, where the first two qubits are in state $\ket{01}$ ($\ket{q_2^cq_1^c} = \ket{01}$). Its implementation requires four $C^{(2)}R^\dagger_{\ell+1}$ gates. 

\paragraph{Matrix $D_1^{\mathbb{Z}_8}$} Similarly, $D_1^{\mathbb{Z}_8}$ implements $\tilde{\Omega}^\dagger$ on the $\mathbb{Z}_8$ position register, conditioned on the coin states $\{\ket{110}, \ket{111}\}$, where the first two qubits are in state $\ket{11}$ ($\ket{q_2^cq_1^c} = \ket{11}$). This requires three $C^{(2)}R^\dagger_{\ell+1}$ gates. 

\subsubsection{Gate Complexity and CNOT Scaling}

Table~\ref{tab:gate_complexity_2D} summarizes the gate complexity and upper-bound CNOT cost for the naive and decomposed implementations of $\Sigma
$ for the $\mathbb{Z}_{16} \times \mathbb{Z}_8$ torus graph. Following the same CNOT scaling established in Section~\ref{sec:upper_bound_CNOT_cost_1D}, we apply the linear $16n - 24$ cost model from Rosa et al.~\cite{rosa_optimizing_2025} uniformly across both implementations.

Because the block-diagonal structure of $\Sigma$ decomposes the 2D shift operation into independent 1D phase-shift primitives acting on the $x$ and $y$ position registers, the upper-bound CNOT cost is additive across dimensions and the scaling analysis of Section~\ref{sec:1D_results} carries over directly to each block. However, the efficiency crossover threshold in $k$ is expected to be lower in the 2D setting than in the 1D case, as the higher minimum control degree imposed by the multi-dimensional coin register raises the baseline overhead of the decomposed implementation, bringing the two methods to parity at a smaller value of $k$.

\begin{table}[h]
\centering
\begin{tabular}{lcccc|c}
\hline
Implementation 
& $C^{(3)}R$ 
& $C^{(2)}R$ 
& $C^{(1)}R$ 
& Single-qubit $R$
& CNOT (upper bound) \\
\hline
Naive 
& 28 
& 0
& 0
& 0
& $1120/1_0$\\
Decomposed
& 5
& 12
& 7
& 0
& $502/1_0$\\
\hline
\end{tabular}
\caption{Gate complexity comparison for the naive and decomposed implementations of $\Sigma$ for the $\mathbb{Z}_{16} \times \mathbb{Z}_8$ Cayley graph with generating set $\mathcal{S}=\{(\pm1,0),(\pm2,0),(0,\pm1),(0,\pm2)\}$. CNOT upper bounds are computed using the $16n-24
$ linear scaling of Rosa et al.~\cite{rosa_optimizing_2025}, where $1_0$ denotes the single auxiliary qubit required to maintain linear scaling}
\label{tab:gate_complexity_2D}
\end{table}

\section{Conclusion}
\label{Conclusion}
We have presented a systematic, multi-stage decomposition framework for the shift operator of discrete-time quantum walks on Cayley graphs, generalizing the Boundary QFT scheme of Razzoli et al. to 1D Cayley graphs with inverse-closed and non-inverse-closed generating sets, as well as to $d$-dimensional torus graphs. By hierarchically factorizing the QFT-diagonalized shift operator into structured block components, the method progressively reduces the control degree of the required rotation gates. This redistribution of control logic from high-degree multi-qubit operations to collections of lower-degree controlled gates yields a substantial reduction in upper-bound CNOT cost, as demonstrated analytically and benchmarked numerically across a range of graph parameters.

The gate complexity analysis establishes that the decomposed implementation is strictly advantageous within the regime $k \leq 64$ for inverse-closed graphs and $k \leq 16$ for non-inverse-closed graphs, where $k$ denotes the degree of the generating set. Within these regimes, the CNOT savings are largely insensitive to the system size $N$, confirming that $k$, and not the number of nodes, is the primary resource bottleneck for the shift operator. The concrete circuit constructions provided for the 8-Cayley graph and the $\mathbb{Z}_{16} \times \mathbb{Z}_8$ torus graph illustrate the practical applicability of the framework, and the modular structure of the 2D decomposition extends naturally to arbitrary dimension.

Several directions remain open and are of direct relevance to practical DTQW circuit implementations. A natural extension is to generalize the decomposition to Cayley graphs whose generating sets are not consecutive integer shifts, as many physically and algorithmically motivated graphs, including expander graphs and graphs arising in quantum-enhanced optimization, have irregular connectivity.

Future research could explore the extension of this decomposition strategy to Cayley graphs with arbitrary connection, where the generators do not form a consecutive set. Other types of graphs, such as non-Abelian Cayley graphs or more complex, non-uniform graph structures can be insightful. Another direction involves integrating these circuits with hardware-aware mapping techniques to further minimize the impact of decoherence on specific quantum processors. Finally, applying these gate-efficient shift operators to higher-level algorithms, such as quantum-enhanced path planning for robotics, remains an area for further investigation.

From a resource estimation perspective, a full fault-tolerant analysis of the proposed circuits would be particularly valuable. The present work quantifies CNOT complexity under a linearized upper bound, but the dominant cost on fault-tolerant architectures is the $T$-gate count arising from the decomposition of the controlled rotation gates $C^{(\alpha)}R_\lambda$ into Clifford+$T$ circuits. Characterizing this cost, and identifying whether the hierarchical structure introduced here propagates favorably into the $T$-gate regime, is an important open question for assessing the long-term utility of the method.

Hardware-aware compilation presents another critical gap. Adapting the decomposition to native gate sets and device topologies could yield further reductions in executable circuit depth beyond what the CNOT upper bound captures. Experimental verification of small instances on current hardware, including characterization of the fidelity gains attributable to reduced circuit, would provide an important empirical complement to the theoretical benchmarks presented here.

\section*{Acknowledgments}
The author would like to thank Hanmeng Zhan for valuable guidance and discussions throughout the course of this work. 

\bibliographystyle{unsrt}  
\bibliography{DTQW}

\end{document}